\documentclass[showpacs,amssymb,twocolumn,floatfix,pre]{revtex4-1}

\usepackage{amssymb,amsfonts}
\usepackage{graphics,graphicx}

\begin{document}

\title{Counter-ions at Charged Walls: Two Dimensional Systems}

\author{Ladislav \v{S}amaj}
\altaffiliation[On leave from ]
{Institute of Physics, Slovak Academy of Sciences, Bratislava, Slovakia}
\email{Ladislav.Samaj@savba.sk}
\author{Emmanuel Trizac}
\affiliation{Universit\'e Paris-Sud, Laboratoire de Physique Th\'eorique et 
Mod\`eles Statistiques, UMR CNRS 8626, 91405 Orsay, France}

\begin{abstract}
We study equilibrium statistical mechanics of classical 
point counter-ions,
formulated on 2D Euclidean space with logarithmic Coulomb interactions 
(infinite number of particles) or on the cylinder surface 
(finite particle numbers), in the vicinity of a single uniformly charged line
(one single double-layer), or between two such lines (interacting double-layers).
The weak-coupling Poisson-Boltzmann theory, which applies
when the coupling constant $\Gamma$ is small, is briefly recapitulated (the coupling constant is defined 
as $\Gamma\equiv \beta e^2$ where $\beta$ is the inverse temperature,
and $e$ the counter-ion charge).
The opposite strong-coupling limit ($\Gamma \to \infty$) 
is treated by using a recent method
based on an exact expansion around the ground-state Wigner crystal of 
counter-ions.
The weak- and strong-coupling theories are compared at intermediary values 
of the coupling constant $\Gamma=2\gamma$ $(\gamma=1,2,3)$, to exact results
derived
within a 1D lattice representation of 2D Coulomb systems in terms of 
anti-commuting field variables. 
The models (density profile, pressure) are solved exactly for any particles 
numbers $N$ at $\Gamma=2$ and up to relatively large finite $N$ at 
$\Gamma=4$ and 6.
For the one-line geometry, the decay of the density profile at asymptotic
distance from the line undergoes a fundamental change with respect to 
the mean-field behavior at $\Gamma=6$. The like-charge attraction regime,
possible in the strong coupling limit but precluded at mean-field level, 
survives for $\Gamma=4$ and 6,
but disappears at $\Gamma=2$.
\end{abstract}



\maketitle

\renewcommand{\theequation}{1.\arabic{equation}}
\setcounter{equation}{0}

\section{Introduction}

Most mesoscopic objects, when dissolved in a polar solvent
such as water, acquire an electric charge through the dissociation 
of functional surface groups \cite{Hunter}. Counter-ions are then released in 
the solution, and form, together with the charged object,
the so-called electric double-layer.
Since the pioneering work of Gouy and Chapman a century ago \cite{GC}, 
the study of these charge density clouds has formed an active line
of research, in particular from a theoretical perspective
\cite{Attard,JPH,Messina09,Boroudjerdi05}. Electric double layers 
are indeed pivotal in affecting single mesoscopic ``particle'' 
properties, together with inter-particle interactions.

The present paper concerns the equilibrium statistical mechanics of
charged particles in the vicinity of charged walls (planar double-layers).
The general problem of mobile ions confined by uniformly charged interfaces 
can be formulated in two ways. 
In the case of ``counter-ions only'', there is just one species of 
equally charged ions neutralizing the surface charge on walls.
In the case of ``added electrolyte'', the system is in contact with an 
infinite reservoir of $\pm$ electrolyte charges.
Here, we shall restrict ourselves to the former class of models
(no salt case). 
The mostly studied 3D geometries are one planar wall with 
counter-ions localized in the complementary half-space and 
two parallel planar walls with counter-ions being localized in between.
The high-temperature (weak-coupling) limit of general Coulomb systems
is described by the Poisson-Boltzmann (PB) mean-field theory
\cite{Andelman06} which basically does not depend on dimension.
Formulating the Coulomb system as a field theory, the PB equation
can be viewed as a first-order term of a systematic expansion in loops
\cite{Attard88,Podgornik88,Podgornik90,Netz00}.

A relevant progress has been made in the last decade 
in the opposite low-temperature (strong-coupling, SC) limit 
\cite{Rouzina96,Grosberg02,Levin02,Moreira00,Netz01,Moreira,Lau1,Lau2,
Boroudjerdi05,Chen06,Santangelo06,Jho08,Dean09,Hatlo10,Kanduc1,Kanduc2}.
Within a field-theoretical treatment \cite{Moreira00,Netz01},
the leading SC behavior stems from a single-particle picture and 
next correction orders correspond to a virial/fugacity expansion
in inverse powers of the coupling constant.
The method requires a renormalization of infrared divergences via 
the electroneutrality condition.
A comparison with the Monte-Carlo simulations \cite{Moreira}
confirmed the adequacy of the leading single-particle theory,
but the predictions for the first correction turns out to be incorrect.
In essence, a virial-like expansion fails, as it also fails
for simple electrolytes (where a reminiscent divergence of 
the second  virial coefficient is indicative of a
non analyticity in the density expansion of the pressure,
and signals that an alternative theoretical
route should be explored \cite{Levin02}).
Very recently, a method based on an exact expansion around
the ground state (the 2D Wigner crystal of counter-ions formed
on the surface of charged walls) was proposed in Ref. \cite{Samaj10}
to overcome such shortcoming.
The expansion is systematic and free of divergences, it provides the correct 
correction to leading SC behaviour, and the obtained results are
in excellent agreement with available data of Monte-Carlo simulations
under strong couplings \cite{Samaj10}.

\begin{figure}[htb]
\begin{center}
\includegraphics[width=0.48\textwidth,clip]{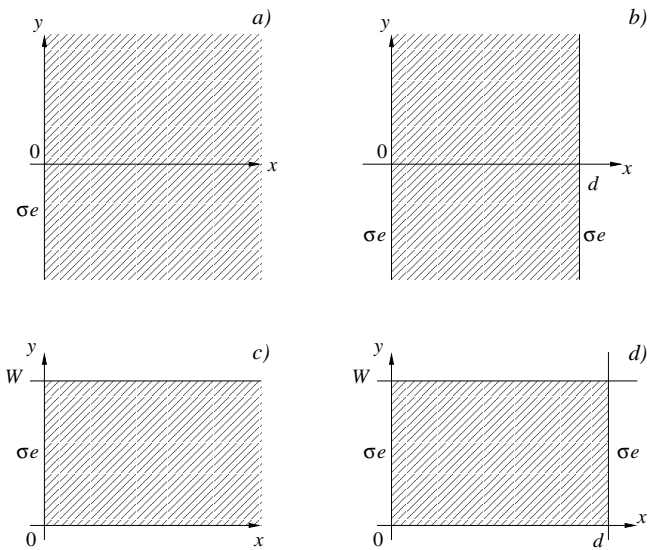}
\caption{Considered geometries: $a)$ one infinite charged line;
$b)$ two parallel charged lines; $c)$ the semi-infinite cylinder 
of circumference $W$; $d)$ finite cylinder. Hatched domains are accessible 
to particles, $\sigma e$ is the uniform charge density of lines.} 
\end{center}
\end{figure}

The majority of previous studies has been restricted to 3D which is 
the dimension of practical interest.
Among notable exceptions are references \cite{Naji05,Burak06},
dealing with 2D Coulomb systems of counter-ions with logarithmic 
pairwise interactions, mainly in the context of Manning condensation. 
Here, we shall concentrate on 2D systems of counter-ions, 
with charged lines on boundary walls,
and Ref. \cite{Dean09} in one dimension.
The homogeneous line charge density is $\sigma e$ ($e$ is the elementary 
charge and $\sigma>0$); point-like counter-ions are for simplicity monovalent 
with charge $-e$.
The studied geometries are pictured in Fig. 1; white walls
are impenetrable to particles, domains accessible to particles are hatched.
The geometries of one (infinite) charged line and two parallel charged lines
at distance $d$ are presented in Figs. 1a and 1b, respectively.
The semi-infinite and finite cylinder surface area of circumference $W$ with 
charged circle boundaries are pictured in Figs. 1c and 1d, respectively.
These cylindric geometries will enable us to mimic the former ones, 
obtained as the infinite-particle-number limit $W\sigma\to\infty$, for 
finite numbers of counter-ions.  
The walls and the confining domains are assumed to possess, for simplicity,
the same (vacuum) dielectric constant $\epsilon=1$, so that there are no 
image charges.
The relevant dimensionless coupling constant is $\Gamma=\beta e^2$, 
where $\beta$ is the inverse temperature.

Interactions of counter-ions with each other and with charged
surfaces are determined by 2D electrostatics defined as follows. 
In $\nu$ spatial dimensions, the electrostatic potential $v$ at
a point ${\bf r}\in {\rm R}^{\nu}$, induced by a unit charge at
the origin ${\bf 0}$, is the solution of Poisson's equation
\begin{equation} \label{1.1}
\Delta v({\bf r}) = - s_{\nu} \delta({\bf r}) ,
\end{equation}
where $s_{\nu}$ is the surface area of the $\nu$-dimensional unit sphere;
$s_2=2\pi$, $s_3=4\pi$, etc.
This definition of the $\nu$-dimensional Coulomb potential maintains
generic properties --such as screening sum rules-- of ``real''
3D Coulomb systems with the interaction potential $v({\bf r}) = 1/r$,
${\bf r}\in {\rm R}^3$ and $r=\vert {\bf r}\vert$.
In an infinite 2D Euclidean space, the solution of (\ref{1.1}), subject to 
the boundary condition $\nabla v({\bf r})\to 0$ as $r\to\infty$, reads
\begin{equation} \label{1.2}
v({\bf r}) = - \ln\left( \frac{r}{L} \right) , \qquad {\bf r}\in {\rm R}^2,
\end{equation} 
where the free-length scale $L$ will be set for simplicity to unity.
Such potential is created by infinitely long charged lines in 3D
which are perpendicular to the given 2D plane; the corresponding systems
occurring in the nature are the so-called polyelectrolytes.
The Coulomb potential (\ref{1.2}) is used for the geometries in 
Figs. 1a and 1b.
For cylindric geometries in Figs. 1c and 1d, the requirement of periodicity
along the $y$-axis, with period $W$, leads to the Coulomb potential 
\cite{Choquard81}
\begin{equation} \label{1.3}
v({\bf r}) = - \ln\left\vert 2\sinh\left( \frac{\pi z}{W}\right) \right\vert , 
\end{equation} 
where the complex notation $z=x+{\rm i}y$ is used for point ${\bf r}=(x,y)$.
At small distances $r\ll W$, this potential behaves like
the logarithmic one (\ref{1.2}) with $L=W/(2\pi)$. 
At large distances along the cylinder $x\gg W$, it behaves like 
the 1D Coulomb potential $-(\pi/W)\vert x\vert$.

Maintaining basic features of physical phenomena, 2D models with one type
of charges have interesting advantages in comparison with 3D ones: 
They are  
less laborious and some of the concepts can be often verified by 
explicit calculations.
2D Coulomb systems are even exactly solvable at a special 
coupling $\Gamma=2$, in infinite space as well as in inhomogeneous 
semi-infinite or finite domains; for a review, see \cite{Jancovici92}. 
Such exactly solvable models can serve as an adequacy test of
weak- and strong-coupling series expansions for finite temperatures.
It was shown in Ref. \cite{Samaj95} that for the sequence of  couplings 
$\Gamma=2\gamma$ $(\gamma=1,2,3,\ldots)$ statistical averages in 
2D Coulomb models can be treated within a 1D lattice theory of
interacting fields of anti-commuting (Grassmann) variables.
This 1D representation enables one to treat exactly one-component Coulomb 
systems with relatively large numbers of particles also for $\Gamma=4,6$.
The technique of anti-commuting variables was applied to the general
problem of integrability of the 2D jellium \cite{Samaj04a} and
to the translational symmetry breaking of the jellium formulated
on the cylinder surface \cite{Samaj04b}.

The present work was motivated by the need to have a control over weak- and
strong-coupling theories.
The exact results for density profiles and pressures at intermediary values 
of $\Gamma$ provide valuable tests of weak- and strong-coupling theories. 
Among the results obtained in this paper, two deserve special attention.
For the one-line geometry, the asymptotic decay of the density profile
from the line undergoes a fundamental change from the mean-field behavior 
at $\Gamma=6$. 
This means that the long-distance predictions of the PB theory have
a restricted validity, which is relevant from the point of view of
the renormalized-charge concept (for a review, see \cite{Levin02}).
This also invalidates the hypothesis that a strongly coupled double-layer, 
at large distances, behaves in a (suitably renormalized)
mean-field fashion \cite{Shklovskii,Levin09}.
For two-line geometry, there is evidence about attraction between 
like-charged lines at relatively small couplings $\Gamma=4,6$.
The attraction is observed even in the $N=2$ particle systems.

The paper is organized as follows. For completeness,
it starts with a short recapitulation of the weak-coupling
PB theory for the straight-line(s) geometries in Figs. 1a and 1b. 
The models are subsequently studied in the SC limit, by using 
the new method \cite{Samaj10}, in Sec. 3.  
Intermediary values of the coupling constant $\Gamma$ are investigated
within the technique of 1D Grassmann variables in Sec. 4.
The models, formulated on the cylinder surface in Figs. 1c and 1d, 
are solved exactly (density profile, pressure) for any particles 
numbers $N$ at $\Gamma=2$ and for (relatively large) finite $N$ at 
$\Gamma=4,6$.
The results are compared with those obtained in the weak- and 
strong-coupling limits. Conclusion are finally drawn in section 5.
    
\renewcommand{\theequation}{2.\arabic{equation}}
\setcounter{equation}{0}

\section{Weak-coupling limit} 
We begin with a brief reminder of the weak-coupling PB theory,
adapted for the geometries pictured in Figs. 1a and 1b. 
More details can be found in e.g. \cite{Andelman06}.

\subsection{Single charged line}
We first consider the case of a single infinite line at $x=0$,
carrying positive charge density $\sigma e$ (Fig. 1a).
Let the density of counter-ions in the half-plane $x>0$ be denoted
by $n(x)$; the corresponding charge density is $\rho(x)=-e n(x)$.
The contact theorem for planar wall surfaces
\cite{Henderson,Carnie81,Wennerstrom82} relates the total contact density 
of particles to the surface charge density on the wall and 
the bulk pressure of the fluid $P$.
For 2D systems of identical particles, it reads
\begin{equation} \label{2.1}
\beta P = n(0) - \pi \Gamma \sigma^2 ,
\end{equation}
where $\Gamma=\beta e^2$ is the coupling constant.
Since in the present case of a single isolated double-layer,
the pressure vanishes ($P=0$), we have $n(0) = \pi \Gamma \sigma^2$.

The induced average electrostatic potential $\phi(x)$ 
is determined by the Poisson equation
\begin{equation} \label{2.2}
\frac{{\rm d}^2 \phi(x)}{{\rm d}x^2} = - 2\pi \rho(x) .
\end{equation}
The condition of electroneutrality 
$\sigma e + \int_0^{\infty} {\rm d}x\, \rho(x) = 0$ is equivalent to
the boundary condition
\begin{equation} \label{2.3} 
- \frac{{\rm d}\phi(x)}{{\rm d}x}\Big\vert_{x=0} = 2\pi\sigma e .
\end{equation}
Within the mean-field approach, valid for $\Gamma\to 0$
(``high temperatures''),
the average particle density is approximated by replacing the potential of 
 mean force by the average electrostatic potential, 
$n(x) = n_0 \exp[\beta e \phi(x)]$.
Equation (\ref{2.2}) then reduces to the nonlinear PB equation
\begin{equation} \label{2.4}
\frac{{\rm d}^2 \phi(x)}{{\rm d}x^2} = 2\pi e n_0 \exp[\beta e \phi(x)] .
\end{equation}
This second-order differential equation, supplemented by the boundary
condition (\ref{2.3}), can be integrated explicitly. 
The solution for the counter-ion density profile reads
\begin{equation} \label{2.5}
n(x) = \frac{1}{\pi\Gamma} \frac{1}{(x+\mu)^2} .
\end{equation}
Here, $\mu=1/(\pi\Gamma\sigma)$ is the Gouy-Chapman length, i.e. 
the distance from the charged wall at which an isolated counter-ion has 
potential energy equal to thermal energy $k_{\rm B}T=1/\beta$.
In what follows, all lengths will be expressed in units of $\mu$, 
$\tilde{x}=x/\mu$.
Note that at asymptotically large distances from the wall 
$\tilde{x}\to\infty$, the density profile (\ref{2.5}) does not
depend on the line charge density magnitude $\sigma e$, 
$n(x)\sim 1/(\pi\Gamma x^2)$.
We shall see that this interesting phenomenon takes place
also at finite temperature ($\Gamma=2$ and $\Gamma=4$), while
another asymptotic decay seems to hold at $\Gamma=6$.

The density profile will be considered also in the rescaled 
form $\tilde{n}(\tilde{x}) = n(\mu\tilde{x})/(\pi\Gamma\sigma^2)$.
Thus, the solution (\ref{2.5}) can be expressed as
\begin{equation} \label{2.6}
\tilde{n}(\tilde{x}) = \frac{1}{(1+\tilde{x})^2}
\mathop{\sim}_{\tilde{x}\to\infty} \frac{1}{\tilde{x}^2} .
\end{equation}
This density profile satisfies both the electroneutrality condition 
(\ref{2.3}), written in the rescaled form as
\begin{equation} \label{2.7}
\int_0^{\infty} {\rm d}\tilde{x}\, \tilde{n}(\tilde{x}) = 1 ,
\end{equation}
and the contact theorem (\ref{2.1}) with $P=0$, written as
\begin{equation} \label{2.8}
\tilde{n}(0) = 1 .
\end{equation}
It is worth emphasising that as an approximate approach,
PB could {\it a priori} violate the contact theorem. 
This is not the case, as follows from arguments that can
be found in Ref. \cite{TT03}.

\subsection{Two charged lines}
The system of two lines at distance $d$ in Fig. 1b has the $x\to d-x$ 
symmetry, so it is sufficient to consider the interval 
$x\in[ 0,d/2]$.
Because of the mentioned symmetry, the electrostatic potential
satisfies the additional boundary condition at the mid-point $x=d/2$
\begin{equation} \label{2.9}
\frac{{\rm d}\phi(x)}{{\rm d}x}\Big\vert_{x=d/2} = 0 .
\end{equation}

Let us denote by $n_m$  the value of the particle density
at $x=d/2$.
The PB equation (\ref{2.2}), supplemented with the boundary condition 
(\ref{2.9}), can be solved explicitly:
\begin{eqnarray} 
\beta e \phi(x) & = & - \ln \cos^2\left[ K\left(x-\frac{d}{2} \right) 
\right] , \nonumber \\
n(x) & = & \frac{n_m}{\cos^2[K(x-d/2)]} , \label{2.10}
\end{eqnarray}
where the inverse length $K$ is related to $n_m$ via
\begin{equation} \label{2.11}
K^2 = \pi \Gamma n_m .
\end{equation}
The boundary condition at $x=0$ (\ref{2.3}) implies the following
transcendental equation for the real $K$
\begin{equation} \label{2.12}
(K d) \tan(Kd/2) = \frac{d}{\mu} \equiv \tilde{d} .
\end{equation}
According to the contact theorem (\ref{2.1}), the (rescaled) pressure
between the charged lines is given by
\begin{equation} \label{2.13} 
\tilde{P} \equiv \frac{\beta P}{\pi\Gamma\sigma^2} 
= (\mu K)^2 .
\end{equation}
Note that in the PB approximation, 
the pressure between two equivalently charged lines
is always positive. This repulsive behaviour is in agreement 
with general results \cite{ET2000}.

Two limiting cases of the dimensionless distance between 
the lines $\tilde{d}$ are of special interest.
For small distances $\tilde{d}\to 0$, we have
\begin{equation} \label{2.14}
\tilde{P} = \frac{2}{\tilde{d}} - \frac{1}{3} + \frac{2}{45} \tilde{d}
+ {\cal O}(\tilde{d}^2) ,
\end{equation}
while for large distances $\tilde{d}\to\infty$, $Kd\to\pi$ and
\begin{equation} \label{2.15}
\beta P \sim \frac{\pi}{\Gamma} \frac{1}{d^2} .
\end{equation}
We see that the asymptotic decay of $P$ to zero does not depend on $\sigma$; 
this behavior is analogous to that of the density profile in 
the one-line problem. Note that $d$ and $1/\sigma$ 
(or equivalently, the Gouy length)
being the two relevant
length scales in the problem, an algebraic decay of $P$ in $1/d^2$ 
at large distances cannot involve any $\sigma$ dependence,
for dimensional reasons.
A similar remark holds for the density profile of the single 
plate problem, Eq. (\ref{2.5}).

\renewcommand{\theequation}{3.\arabic{equation}}
\setcounter{equation}{0}

\section{Strong-coupling theory}

\subsection{Single charged line}
At zero temperature, i.e. in the strong coupling (SC) limit $\Gamma\to\infty$, 
the counter-ions collapse on the charged line, see Fig. 2a.
They create a 1D Wigner crystal, with vertices 
${\bf R}_j=(0,j/\sigma)$ $(j=0,\pm 1,\pm 2,\ldots)$; the nearest-neighbor 
distance $a=1/\sigma$ ensures the local neutrality on the line.
It is essential here to bear in mind 
that the SC limit corresponds to the regime in which
$a$ is much larger than the characteristic Gouy distance $\mu$ between
the counter-ions and the charged line \cite{Rouzina96},
$\tilde{a}\equiv a/\mu\propto \Gamma\to\infty$.
In the asymptotic SC limit $\Gamma\to\infty$, each vertex ${\bf R}_j$
is occupied by a counter-ion $j$.
The ground-state energy of the discrete particle system plus the 
homogeneous background charge density will be denoted by $E_0$.
For $\Gamma$ large but finite, the fluctuations of counter-ions
around their vertex positions become important. 
Our approach consists in a systematic account of these fluctuations,
and provides an expansion in inverse powers of $\Gamma$. 
It is the exact adaptation to the 2D case of the procedure discussed
in \cite{Samaj10} for three dimensional systems. It differs
from the approach of Netz and collaborators 
\cite{Boroudjerdi05,Netz01,Moreira} in that the latter 
virial-like procedure, although capturing the leading order
term in the expansion, leads to divergences in the corrections.
As a consequence, the corrections to the leading order behaviour
are not of the order in the coupling constant that is predicted
in \cite{Boroudjerdi05,Netz01,Moreira}, see \cite{Samaj10}.
On the other hand, our expansion, that purports to capture the
very same phenomena as Refs \cite{Boroudjerdi05,Netz01,Moreira}, 
is free of divergences, and reveals that 
the corrections to the leading order are much larger than
obtained in \cite{Boroudjerdi05,Netz01,Moreira}. This is
confirmed by the Monte Carlo simulations, that are, for three
dimensional systems, in complete agreement with the predictions
of \cite{Samaj10}. Our strong coupling expansion method
also differs from that put forward in refs \cite{Lau1,Lau2}
by Lau, Pincus, and collaborators. These authors impose that
the counter-ions stick to the charged interfaces, so that the
counter-ions
degrees of freedom are ``in-plane'' only (along the interface). 
This precludes the possibility
to study the phenomenon of like-charge attraction at small distances,
see below. Indeed, it is essential to include in the
analysis the excitations where the counter-ions can unbind from the 
charged interface (displacements perpendicular to the interface).
Consequently, the results of Refs. \cite{Lau1,Lau2} should
be viewed as a large distance expansion, and in this respect, 
complementary to our short distance analysis.

\begin{figure}[htb]
\begin{center}
\includegraphics[width=0.48\textwidth,clip]{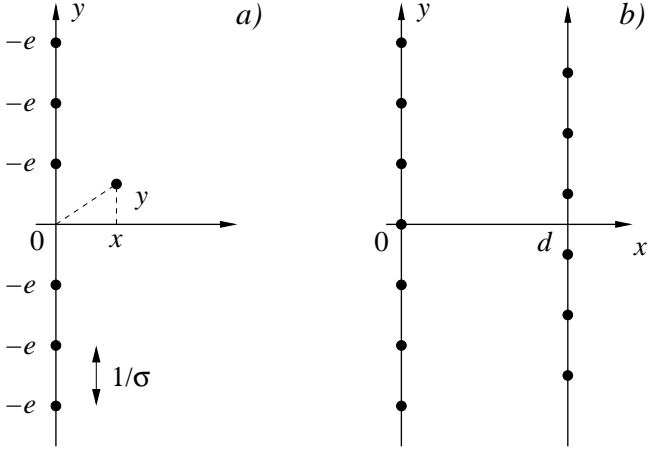}
\caption{Ground-state Wigner crystal of counter-ions (shown by the black dots) 
in  $a)$ one-line geometry; $b)$ two-lines geometry. Figure a) also introduces
the notation $(x,y)$ for the position shift of a tagged counter-ion,
that will be used in our analytical calculation.} 
\end{center}
\end{figure}

As depicted in Fig. 2.a, let us first shift one of the particles, say $j=0$, from its lattice
position ${\bf R}_0=(0,0)$ by a small vector $\delta{\bf R}_0=(x>0,y)$,
$\delta R_0\ll a$.
The corresponding change in the total energy 
$\delta E(x,y) = E(x,y)-E_0\ge 0$ consists of two contributions.
The first one is due to the interaction of the shifted counter-ion
with the uniform line charge density:
\begin{equation} \label{3.1}
\delta E^{(1)}(x) = \pi e^2 \sigma x .
\end{equation}
The second contribution $\delta E^{(2)}(x,y)$ comes from the interaction of 
the shifted particle with all other ions on the 1D Wigner crystal.
For the special case of the $y=0$ shift, we find
\begin{eqnarray} 
\delta E^{(2)}(x,0) & = & - e^2 \sum_{j=-\infty\atop (j\ne 0)}^{\infty}
\left[ \ln \sqrt{x^2+\left( \frac{j}{\sigma}\right)^2}
- \ln\left( \frac{\vert j\vert}{\sigma}\right) \right] \nonumber \\
& = & - e^2 \ln \prod_{j=1}^{\infty} 
\left[ 1 + \left( \frac{\sigma x}{j} \right)^2 \right] \nonumber \\
& = & - e^2 \ln \left[ \frac{\sinh(\pi\sigma x)}{\pi\sigma x} \right] . 
\label{3.2} 
\end{eqnarray} 
This function has a small-$x$ expansion of the form
\begin{equation} \label{3.3}
\delta E^{(2)}(x,0) = - \frac{e^2}{6} (\pi\sigma x)^2
+ \frac{e^2}{180} (\pi\sigma x)^4 + {\cal O}(x^6) .
\end{equation}
Note that the interaction with counter-ions does not imply any contribution
linear in $x$. The corresponding electric field indeed vanishes,
by symmetry.
The negative sign of the leading $x^2$ term does not represent any problem:
The sum of $\delta E^{(2)}(x,0)$ and the linear term (\ref{3.1}) is 
a monotonously increasing function of $x$, as it should be.
For the special case of the $x=0$ shift, we find
\begin{eqnarray}
\delta E^{(2)}(0,y) & = & - e^2 \ln \prod_{j=1}^{\infty} 
\left[ 1 - \left( \frac{\sigma y}{j} \right)^2 \right] \nonumber \\
& = & - e^2 \ln \left[ \frac{\sin(\pi\sigma y)}{\pi\sigma y} \right] 
\nonumber \\ & = & 
\frac{e^2}{6} (\pi\sigma y)^2 + \frac{e^2}{180} (\pi\sigma y)^4 
+ {\cal O}(y^6) . \label{3.4}
\end{eqnarray}
The calculation of the whole function $\delta E^{(2)}(x,y)$ for the particle 
shift simultaneously along both directions is complicated.
For our purpose, it is sufficient to derive its expansion in $x,y$ up
to harmonic terms:
\begin{eqnarray}
\delta E^{(2)}(x,y) & = & - \frac{e^2}{2} \sum_{j=1}^{\infty} \ln
\Bigg\{ \left[ 1 + \left( \frac{\sigma x}{j} \right)^2 +
\left( \frac{\sigma y}{j} \right)^2 \right]^2 \nonumber \\
& & - 4 \left( \frac{\sigma y}{j} \right)^2 \Bigg\} 
\sim \frac{e^2(\pi\sigma)^2}{6} (y^2-x^2) .  \label{3.5}
\end{eqnarray}
where use was made of $\sum_{j=1}^{\infty} 1/j^2 = \pi^2/6$ 
\cite{Gradshteyn}.
Note the absence of the mixed harmonic term $xy$.
The total energy change, up to harmonic terms and in dimensionless form, 
is finally given by
\begin{equation} \label{3.6}
-\beta \delta E(x,y) \sim -\tilde{x} + 
\frac{1}{6\Gamma}(\tilde{x}^2-\tilde{y}^2) .
\end{equation}
This formula reveals a relationship between the order of the expansion
of $-\beta E(x,y)$ in the dimensionless lengths $\tilde{x},\tilde{y}$
and its SC expansion in $1/\Gamma$.
The linear term $-\tilde{x}$ is the only one which does not vanish
in the limit $\Gamma\to\infty$.
This leading term reflects the single-particle character of the
system close to zero temperature: Each particle behaves independently
from the others, exposed only to the field of the line charge density.
A similar conclusion is reached within the original approach of
Netz and collaborators \cite{Boroudjerdi05,Moreira,Netz01}.
Terms of the $p$th order in $x,y$ have prefactors proportional to
$\Gamma\sigma^p$, i.e. passing to $\tilde{x},\tilde{y}$ they become
of the SC order $1/\Gamma^{p-1}$.  
This is why the first $1/\Gamma$ correction to the single-particle
regime is determined exclusively by the terms harmonic in space.

Let us now consider a shift of all particles from their lattice positions 
${\bf R}_j$ $(j=0,\pm 1,\pm 2,\ldots)$ by small vectors 
$\delta{\bf R}_j=(x_j,y_j)$.
To determine the corresponding increase in the total energy
$\delta E(\{x_j,y_j\})$, we proceed as above and obtain
\begin{eqnarray}
-\beta\delta E(\{ x_j,y_j\}) & \sim & -\sum_j \tilde{x}_j
+\frac{1}{2\pi^2\Gamma} \sum_{j<k} \frac{(\tilde{x}_j-\tilde{x}_k)^2}{(j-k)^2} 
\nonumber \\ & & - \frac{1}{2\pi^2\Gamma} 
\sum_{j<k} \frac{(\tilde{y}_j-\tilde{y}_k)^2}{(j-k)^2} . \label{3.7}
\end{eqnarray}

The $x$-dependent density profile of counter-ions is defined as the average
$n(x) = \langle\sum_{j=1}^N \delta({\bf r}-{\bf r}_j)\rangle$.
The mean-value calculation with the energy (\ref{3.7}) is simplified
in two ways.
Firstly, since the $y$-coordinates do not mix with $x$-coordinates,
they can be neglected in the considered order.
Secondly, since all (identical) particles are exposed to the same
one-body potential of the charged line, a summation over particle
degrees of freedom can be represented by just one auxiliary coordinate.
We find
\begin{eqnarray}
\tilde{n}(\tilde{x}) & = & 
C {\rm e}^{-\tilde{x}} \int_0^{\infty} {\rm d}\tilde{x}'\,
{\rm e}^{-\tilde{x}'} \left[ 1 + \frac{1}{2\pi^2\Gamma} 
\sum_{j=-\infty}^{\infty}\frac{(\tilde{x}-\tilde{x}')^2}{j^2} \right]
\nonumber \\ & & + {\cal O}\left( \frac{1}{\Gamma^2} \right) , \label{3.8} 
\end{eqnarray}
where $C$ is determined by the normalization condition (\ref{2.7}).
Simple algebra yields
\begin{equation} \label{3.9}
\tilde{n}(\tilde{x}) = {\rm e}^{-\tilde{x}} \left[ 1 + \frac{1}{3\Gamma} 
\left( \frac{\tilde{x}^2}{2}-\tilde{x} \right) \right]
+ {\cal O}\left( \frac{1}{\Gamma^2} \right) .
\end{equation}
The contact theorem (\ref{2.8}) is fulfilled by this profile.

\subsection{Two charged lines}
In the problem of two lines at distance $d$, carrying the same charge 
density $\sigma e$, the electric field between the lines vanishes.
At zero temperature $\Gamma\to\infty$, the counter-ions collapse on
the lines, see Fig. 2b.
The Wigner crystal is thus composed of two one-dimensional arrays of sites
with the lattice constant $a=1/\sigma$, shifted with respect to one another
by half-period $a/2$.
The vertices of the Wigner lattice will be denoted as
${\bf R}_j^{(0)}=(0,j/\sigma)$ $(j=0,\pm 1,\pm 2,\ldots)$ if they belong
to the line at $x=0$ and as
${\bf R}_j^{(d)}=(d,j/\sigma)$ $(j=\pm \frac{1}{2},\pm \frac{3}{2},\ldots)$ 
if they belong to the line at $x=d$.
The SC regime of large $\Gamma$ lends itself to analytic progress
when the inequality $d\ll a$, or equivalently
\begin{equation} \label{3.10}
d\sigma \ll 1 
\end{equation}
is obeyed.  
As before, the shifts of particles from their Wigner positions along 
the $y$ direction have no effects on statistical averages in 
the leading SC order and in the first $1/\Gamma$ correction,
so we shall not consider them.
Let first one of the particles, say the one localized on the $x=0$ line 
at ${\bf R}_0^{(0)}=(0,0)$, be shifted along the $x$-direction by 
a small amount $x>0$.
The corresponding energy change, up to harmonic terms, is given by
\begin{equation} \label{3.11}
-\beta\delta E(x) \sim \frac{1}{6\Gamma}\tilde{x}^2
- \frac{1}{2\Gamma} \left[ \tilde{d}^2-(\tilde{d}-\tilde{x})^2\right] ,  
\end{equation}
where we used the formula $\sum_{j=\frac{1}{2},\frac{3}{2},\ldots}1/j^2 = \pi^2/2$.
When all particles are shifted in the $x$ direction within the area
limited by the two lines, ${\bf R}_j^{(0)}\to (x_j^{(0)},j/\sigma)$
$(j=0,\pm 1,\pm 2,\ldots)$ and ${\bf R}_j^{(d)}\to (x_j^{(d)},j/\sigma)$
$(j=\pm \frac{1}{2},\pm \frac{2}{2},\ldots)$, the energy change is    
given by
\begin{eqnarray}
-\beta\delta E(\{ x_j\}) & \sim & \frac{1}{2\pi^2\Gamma} 
\sum_{j<k} \frac{[\tilde{x}_j^{(0)}-\tilde{x}_k^{(0)}]^2}{(j-k)^2} \nonumber \\
& & + \frac{1}{2\pi^2\Gamma} 
\sum_{j<k} \frac{[\tilde{x}_j^{(d)}-\tilde{x}_k^{(d)}]^2}{(j-k)^2} 
\nonumber \\ & & - \frac{1}{2\pi^2\Gamma} \sum_{j,k} 
\frac{\tilde{d}^2-[\tilde{x}_j^{(0)}-\tilde{x}_k^{(d)}]^2}{(j-k)^2} . 
\phantom{aaa} \label{3.12} 
\end{eqnarray}

The density of counter-ions at $x$ $(0\le x\le d)$ is calculated with 
the energy (\ref{3.12}):
\begin{eqnarray}
\tilde{n}(\tilde{x}) & = & C \int_0^{\tilde{d}} {\rm d}\tilde{x}'\,
\left[ 1 + \frac{1}{2\pi^2\Gamma} \sum_{j=\pm 1,\pm 2,\ldots}
\frac{(\tilde{x}-\tilde{x}')^2}{j^2} \right. \nonumber \\
& & \left. +\frac{1}{2\pi^2\Gamma} \sum_{j=\pm \frac{1}{2},\pm \frac{3}{2},\ldots}
\frac{(\tilde{x}-\tilde{x}')^2}{j^2} \right] 
+ {\cal O}\left(\frac{1}{\Gamma^2}\right) , \nonumber \\ & & \label{3.13} 
\end{eqnarray}
where $C$ is determined by the normalization condition
\begin{equation} \label{3.14}
\int_0^{\tilde{d}} {\rm d}\tilde{x}\, \tilde{n}(\tilde{x}) = 2 .
\end{equation} 
After simple algebra, we arrive at
\begin{equation} \label{3.15}
\tilde{n}(\tilde{x}) = \frac{2}{\tilde{d}} \left\{
1 + \frac{2}{3\Gamma} \left[ \left( \tilde{x}-\frac{\tilde{d}}{2} \right)^2
- \frac{\tilde{d}^2}{12} \right] \right\} 
+ {\cal O}\left(\frac{1}{\Gamma^2}\right) .
\end{equation}
This expression has the needed $x\to d-x$ symmetry.

To derive the pressure between the lines, we apply the contact 
theorem (\ref{2.1}) to obtain, in the SC $\Gamma\to\infty$ limit,
\begin{equation} \label{3.16}
\tilde{P} = \frac{2}{\tilde{d}} - 1 + \frac{2\tilde{d}}{9\Gamma}
+ {\cal O} \left( \frac{\tilde{d}^{\,2}}{\Gamma^2} \right)  .
\end{equation}
Note that the small-$\tilde{d}$ pressure expression (\ref{2.14}), 
obtained in the weak-coupling limit, and (\ref{3.16}) obtained in 
the strong-coupling limit, coincide only in the leading $2/\tilde{d}$ term.
The sub-leading constant terms differ. It should be kept in mind here 
that the present expansion makes sense provided Eq. (\ref{3.10})
is fulfilled, which means that $\tilde d \ll \Gamma$. 
Given that (\ref{3.16}) exhibits an attractive regime, for large enough 
$\Gamma$, for values of $\tilde d$ slightly above 2, 
it can be concluded that Eq.  (\ref{3.16}) is able to capture
the possibility of like-charge attraction, a phenomenon 
first reported some 30 years ago \cite{Gulbrand84,Kjellander84,Kekicheff93}.
It is also interesting to comment on the mean-field failure
to capture such an effect, in spite of a small $d$ expansion 
of the pressure, Eq. (\ref{2.14}), that is very close to its SC counterpart
(\ref{3.16}). The requirement for the validity of expansion (\ref{2.14})
is that $\tilde d \ll 1$, outside the
regime of $\tilde d >6$ where Eq. (\ref{2.14}), 
taken without particular precaution,
would lead to a negative pressure.  The requirement $\tilde d \ll 1$
stands in the strongly confined
regime, where the entropic cost of confining the counter-ions 
is overwhelming, and leads to a dominant repulsive 
pressure $\tilde P \sim 2/\tilde d$.

\begin{figure}[htb]
\begin{center}
\includegraphics[width=0.48\textwidth,clip]{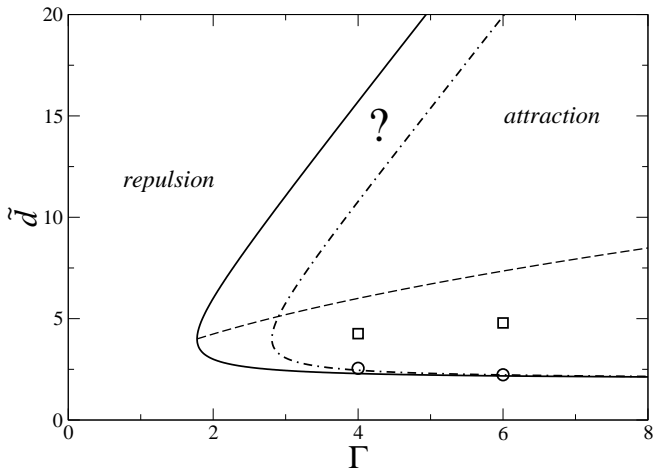}
\caption{Phase boundaries following from the equation of state (\ref{3.16}).
The solid curve corresponds to $\tilde P=0$, and discriminates the repulsive region 
on the left hand side from the attractive one.
The dashed curve corresponds to
the maximum attraction, given by $\partial\tilde{P}/\partial\tilde{d}=0$.
The dashed-dotted line stands for the phase boundary following from the
refined equation of state (\ref{4.47}), where the function of coupling 
$A_N$ is given by its infinite $N$ extrapolation 
(\ref{4.50}). The two circles stand for the exact 
location of the $P=0$ points for $\Gamma=4$ and 6, obtained in section 4.
Likewise, the two squares are the corresponding exact maximum attraction 
points at $\Gamma=4$ and 6 (see text). We stress that such diagrams 
are obtained
from strong coupling in conjunction with short distance
expansions. Consequently, for the range of $\Gamma$ values plotted,
a qualitative picture only can be expected. The question mark is a reminder
of the fact that $\tilde d$ should be much smaller than $\Gamma$ to
allow for our expansion.} 
\end{center}
\end{figure}

Taking the expansion order as indicated in formula (\ref{3.16}), 
the attractive $(P<0)$ and repulsive $(P>0)$ regions in 
the $(\Gamma,\tilde{d})$ plane are split by the solid curve in Fig. 3.
The maximum attraction, given by $\partial\tilde{P}/\partial\tilde{d}=0$, 
is obtained for $\tilde{d}_{\rm max}=3\sqrt{\Gamma}$, see the dashed line.
Open symbols for $\Gamma=4,6$ are the results obtained in the next 
section from an analysis of finite particle numbers. 
It is important here to emphasize that the equation of state 
leading to Fig. 3 is trustworthy for $\tilde d \ll \Gamma$.
As a consequence, the dashed
line showing the locus of maximal attraction becomes asymptotically
exact (since $\tilde d_{\text{max}} /\Gamma \propto \Gamma^{-1/2}$ 
vanishes at large coulombic 
couplings), but 
the upper part of that diagram in Fig 3 is only
indicative, as the question mark indicates. It will in particular
be shown in the next section that the re-entrance of the repulsive 
regime at fixed $\Gamma$, i.e. the fact there exist two distances
$\tilde d$ where the pressure vanishes, does not take place
at $\Gamma=4$ nor at $\Gamma=6$. For such couplings, the first 
zero of $P$ only (with $\tilde d$ slightly above 2) is 
observed. Consequently, while the re-entrance phenomenon is a feature 
backed by Monte Carlo simulations for three dimensional systems
\cite{Moreira}, its existence in the present two dimensional situation 
is still an open question, that is difficult to address with analytical tools.

\renewcommand{\theequation}{4.\arabic{equation}}
\setcounter{equation}{0}

\section{``Intermediary'' fluid regime}

\subsection{General formalism}
We consider $N$ identical particles of charge $-e$, constrained to 
a 2D domain $\Lambda$; points of the domain will be specified by 
the complex coordinates $z=x+{\rm i}y,\bar{z}=x-{\rm i}y$.
The particles interact pair-wisely through the 2D Coulomb potential 
$v(z,z')=-\ln\vert z-z'\vert$ and are exposed to a one-body potential 
$e \psi(z,\bar{z})$ due to the uniform charge density $\sigma e$ on 
the line walls forming the domain boundary $\partial\Lambda$. 
The partition function at coupling $\Gamma$ is defined as 
\begin{equation} \label{4.1}
Z_N = \frac{1}{N!} \int_{\Lambda}\prod_{j=1}^N \left[ {\rm d}^2 z_j\,
w(z_j,\bar{z}_j) \right] \prod_{j<k} \vert z_j-z_k \vert^{\Gamma} ,
\end{equation}
where $w(z,\bar{z}) = \exp[\Gamma \psi(z,\bar{z})]$ is 
the one-body Boltzmann factor.
$\ln Z_N$ is the generator for the particle density in the following sense 
\begin{equation} \label{4.2}
n(z,\bar{z}) = w(z,\bar{z}) \frac{\delta}{\delta w(z,\bar{z})} \ln Z_N .
\end{equation}

For the couplings $\Gamma=2\gamma$ with $\gamma$ a positive integer, 
a 1D Grassmann representation of the 2D partition function (\ref{4.1})  
was derived and further developed in a series of works 
\cite{Samaj95,Samaj04a,Samaj04b}.
Let us introduce on a discrete chain of $N$ sites $j=0,1,\ldots,N-1$ 
two sets of Grassmann variables $\{ \xi_j^{(\alpha)},\psi_j^{(\alpha)}\}$, 
each with $\gamma$ components $\alpha=1,\ldots,\gamma$.
The Grassmann variables satisfy the ordinary anti-commuting algebra
\cite{Berezin66}.
The partition function (\ref{4.1}) is expressible as an integral
over the Grassmann variables in the following way
\begin{eqnarray} 
Z_N(\gamma) & = & \int {\cal D}\psi {\cal D}\xi\, {\rm e}^{S(\xi,\psi)} ,
\nonumber \\ S(\xi,\psi) & = & \sum_{j,k=0}^{\gamma(N-1)} \Xi_j w_{jk} \Psi_k .
\label{4.3}
\end{eqnarray}
Here, ${\cal D}\psi {\cal D}\xi = \prod_{j=0}^{N-1} {\rm d}\psi_j^{(\gamma)}
\ldots {\rm d}\psi_j^{(1)} {\rm d}\xi_j^{(\gamma)} \ldots {\rm d}\xi_j^{(1)}$
and the action $S$ involves pair interactions of composite operators
\begin{eqnarray}
\Xi_j & = & \sum_{j_1,\ldots,j_{\gamma}=0\atop (j_1+\ldots+j_{\gamma}=j)}
\xi_{j_1}^{(1)} \cdots \xi_{j_{\gamma}}^{(\gamma)} , \nonumber \\
\Psi_k & = & \sum_{k_1,\ldots,k_{\gamma}=0\atop (k_1+\ldots+k_{\gamma}=k)}
\xi_{k_1}^{(1)} \cdots \xi_{k_{\gamma}}^{(\gamma)} ,  \label{4.4}
\end{eqnarray}
i.e. the products of all $\gamma$ anti-commuting-field components,
belonging to either $\xi$-set or $\psi$-set, with the fixed sum
of site indices.
The interaction matrix has the elements
\begin{equation} \label{4.5}
w_{jk} = \int_{\Lambda} {\rm d}^2 z\, w(z,\bar{z}) z^j \bar{z}^k ;
\qquad j,k=0,1,\ldots,\gamma(N-1) .
\end{equation}
The representation (\ref{4.3}) provides $Z_N(\gamma)$ as a function of
interaction elements $\{ w_{jk}\}$.
The particle density (\ref{4.2}) is given by
\begin{equation} \label{4.6}
n(z,\bar{z}) = w(z,\bar{z}) \sum_{j,k=0}^{\gamma(N-1)} 
\langle \Xi_j \Psi_k \rangle z^j \bar{z}^k ,
\end{equation}
where the two-correlators
\begin{eqnarray}
\langle \Xi_j \Psi_k \rangle & \equiv & \frac{1}{Z_N(\gamma)}
\int {\cal D}\psi {\cal D}\xi\, {\rm e}^{S(\xi,\psi)} \Xi_j \Psi_k \nonumber \\ 
& = & \frac{\partial}{\partial w_{jk}} \ln Z_N(\gamma) . \label{4.7} 
\end{eqnarray}

The above Grassmann formalism is straightforwardly applicable to the case of 
the cylinder surface with the Coulomb potential (\ref{1.3}).
Due to the periodicity along the $y$ axis, both the one-body Boltzmann
factor $w$ and the particle density $n$ are only $x$-dependent. 
The interaction Boltzmann weight for two particles at the points
$z=x+{\rm i}y$ and $z'=x'+{\rm i}y'$ is expressible as
\begin{eqnarray} 
\left\vert 2 \sinh \frac{\pi(z-z')}{W} \right\vert^{\Gamma}
& = & {\rm e}^{\pi\Gamma(x+x')/W} \nonumber \\ & & \times \left\vert 
{\rm e}^{-2\pi z/W} - {\rm e}^{-2\pi z'/W} \right\vert^{\Gamma} . 
\phantom{aaaa} \label{4.8}
\end{eqnarray}
For each particle of the $N$-particle system, the prefactors from $N-1$ 
pairwise interaction Boltzmann weights and the multiplication by 
the constant $4\pi/W^2$ (which is irrelevant from the point of view of 
the particle density) renormalize the one-body $w(x)$ in the following way
\begin{equation} \label{4.9}
w_{\rm ren}(x) = \frac{4\pi}{W^2} w(x) 
\exp\left[ \frac{2\pi\gamma}{W}(N-1)x \right] . 
\end{equation}
The partition function is again given by (\ref{4.1}), with
the substitutions $w\to w_{\rm ren}$ and $z\to \exp(-2\pi z/W)$.
Due to the orthogonality relation
\begin{equation} \label{4.10}
\int_0^W {\rm d}y\, \exp\left[ \frac{2\pi}{W}{\rm i}(j-k)y \right]
= W \delta_{jk} ,
\end{equation}
the interaction matrix (\ref{4.5}) becomes diagonal,
$w_{jk}=w_j \delta_{jk}$ with
\begin{equation} \label{4.11}
w_j = W \int_{\Lambda} {\rm d}x\, w_{\rm ren}(x)
\exp\left( - \frac{4\pi}{W} j x \right) .
\end{equation}
Due to the ``diagonalized'' form of the partition function
\begin{equation} \label{4.12}
Z_N(\gamma) = \int {\cal D}\psi {\cal D}\xi\, 
\prod_{j=0}^{\gamma(N-1)} \exp\left( \Xi_j w_j \Psi_j \right) , 
\end{equation}
only two-correlators
\begin{equation} \label{4.13}
\langle \Xi_j \Psi_j \rangle = \frac{\partial}{\partial w_j} \ln Z_N(\gamma)
\end{equation}
will be nonzero.
The density is thus given by
\begin{equation} \label{4.14}
n(x) = w_{\rm ren}(x) \sum_{j=0}^{\gamma(N-1)} \langle \Xi_j \Psi_j \rangle
\exp\left( - \frac{4\pi}{W} j x \right) .
\end{equation}
We see that the original problem reduces to finding the explicit dependence 
of $Z_N(\gamma)$ on the set of weights $\{ w_j \}_{j=0}^{\gamma(N-1)}$, 
say by using the anti-commuting integral in (\ref{4.12}).

$Z_N(\gamma)$ can be found trivially for $\gamma=1$ $(\Gamma=2)$ 
when the composite operators are the standard anti-commuting variables 
$\Xi_j=\xi_j, \Psi_j=\psi_j$.
The partition function contains the only term
\begin{equation} \label{4.15}
Z_N(1) = w_0 w_1 \cdots w_{N-1} .
\end{equation}
Consequently,
\begin{equation} \label{4.16}
\langle \Xi_j\Psi_j \rangle = \frac{1}{w_j} \qquad
\mbox{for all $j=0,1,\ldots,N-1$.}
\end{equation} 

In the case of higher integer $\gamma$'s, the number of (always positive) 
terms increases quickly with $N$.
For $N=2$ particles and arbitrary integer $\gamma$, we have
\begin{equation} \label{4.17}
Z_2(\gamma) = \frac{1}{2} \sum_{j=0}^{\gamma} {\gamma \choose j}^2
w_j w_{\gamma-j} .
\end{equation}
For $N=3$ particles, we have
\begin{eqnarray}
Z_3(2) & = & w_0 w_2 w_4 + 2 w_0 w_3^2 + 2 w_1^2 w_4 \nonumber \\
& & + 4 w_1 w_2 w_3 + 6 w_2^3 , \label{4.18} \\
Z_3(3) & = & w_0 w_3 w_6 + 3^2 w_0 w_4 w_5 + 3^2 w_1 w_2 w_6 \nonumber \\
& & + 6^2 w_1 w_3 w_5 + 15^2 w_2 w_3 w_4 , \label{4.19}
\end{eqnarray}
etc. 
To document the number of terms in $Z_N(\gamma)$ we mention that 
when all $w_j=1$ then $Z_N(\gamma)=(\gamma N)!/[(\gamma!)^N N!]$.
The methods for systematic generation of $Z_N(\gamma)$, realized
in practice through computer language {\it Fortran}, 
are summarized in Ref. \cite{Samaj04a}. 
We were able to go up to $N=10$ particles for $\gamma=2$ and
up to $N=9$ particles for $\gamma=3$. 
For the sake of completeness, we mention that 
the number of terms in $Z_N(\gamma)$ is on the order of 
$10^8$ for 10 particles at $\gamma=2$.

For $\gamma$ being an odd positive integer, the composite operators
$\Xi$ and $\Psi$ are products of an odd number of anti-commuting
variables.
This is why they satisfy the usual anti-commutation rules
$\{ \Xi_j,\Xi_k \} = \{ \Psi_j,\Psi_k \} = \{ \Xi_j,\Psi_k \} = 0$
and, in particular, we have $\Xi_j^2 = \Psi_j^2 = 0$.
Each exponential in (\ref{4.12}) is then expanded as
$\exp(\Xi_j w_j\Psi_j) = 1 + \Xi_j w_j \Psi_j$.
We conclude that, for odd $\gamma$, a given $w_j$ can occur
in a summand of $Z_N(\gamma)$ at most once.  
In view of (\ref{4.13}), this property implies the inequality
\begin{equation} \label{4.20}
w_j \langle \Xi_j \Psi_j \rangle \le 1 \qquad
\mbox{for $\gamma=1,3,5,\ldots$.} 
\end{equation}
On the other hand, if $\gamma$ is an even positive integer, the composite 
operators are products of an even number of anti-commuting variables and
therefore commute with each other: 
$[ \Xi_j,\Xi_k ] = [ \Psi_j,\Psi_k ] = [ \Xi_j,\Psi_k ] = 0$.
Higher powers of $w_j$ are then allowed in summands of $Z_N(\gamma)$
and the inequality (\ref{4.20}) has no counterpart for even $\gamma$'s.

\subsection{Cylinder: Single charged line}
We consider the periodic strip of circumference $W$, semi-infinite
in the $x$-direction, $x\in[ 0,\infty]$, see Fig. 1c.
The charge density $\sigma e$ at line $x=0$ is neutralized
by $N=\sigma W$ particles of charge $-e$.
The potential induced by the line charge is $-\pi\sigma e x$,
so that $w(x)=\exp(-\Gamma\pi\sigma x)$.
The renormalized one-body Boltzmann factor (\ref{4.9}) and 
the interaction strengths (\ref{4.11}) take the form
\begin{equation} \label{4.21}
w_{\rm ren}(x) = \frac{4\pi}{W^2} \exp\left( - \frac{2\pi\gamma}{W} x \right) ,
\quad w_j = \frac{1}{j+(\gamma/2)} .
\end{equation}
The particle density (\ref{4.14}) reads
\begin{equation} \label{4.22}
n(x) = \frac{4\pi}{W^2} \sum_{j=0}^{\gamma(N-1)} \langle \Xi_j \Psi_j \rangle
\exp\left[ - \frac{4\pi}{W} \left( j+\frac{\gamma}{2} \right) x \right] .
\end{equation}
For finite $N$ (or, equivalently, finite $W$), the particle density 
exhibits at asymptotically large distances an exponential decay to zero, 
$\lim_{x\to\infty}n(x)\propto \exp(-2\pi\gamma x/W)$.

Our aim is to continualize the formula (\ref{4.22}) in the thermodynamic 
limit $N,W\to \infty$, at the fixed ratio $N/W=\sigma$; this makes
the semi-infinite cylinder surface equivalent to the system of 
the charged straight line in contact with half-space occupied by counter-ions.
For a given finite $N$, we define a set of discrete values 
$f_{j,N}^{(\gamma)}=\gamma w_j \langle \Xi_j\Psi_j \rangle$
with $j=0,1,\ldots,\gamma(N-1)$.
As the continuous variable, we choose $t=j/[\gamma(N-1)]$, taking values 
in the interval $[0,1]$. 
In the continuum limit $N\to\infty$, the set of discrete values 
$f_{j,N}^{(\gamma)}$ tends to a continuous positive function 
\begin{equation} \label{4.23}
f^{(\gamma)}(t) = \lim_{N\to\infty} f_{j,N}^{(\gamma)} , \qquad
\begin{array}{cc}
t = j/[\gamma(N-1)] , \cr
j=0,1,\ldots,\gamma(N-1) .
\end{array}
\end{equation}
The continualization of (\ref{4.22}) results in
\begin{equation} \label{4.24}
\tilde{n}(\tilde{x}) = 2 \int_0^1 {\rm d}t\, t f^{(\gamma)}(t)
\exp(-2 t \tilde{x}) .
\end{equation}
We see that the original problem reduces to the one of finding
the function $f^{(\gamma)}(t)$.
The electroneutrality condition (\ref{2.7}) and the contact theorem 
(\ref{2.8}) hold provided that the function $f^{(\gamma)}(t)$ is constrained by
\begin{equation} \label{4.25}
\int_0^1 {\rm d}t\, f^{(\gamma)}(t) = 1 , \qquad
\int_0^1 {\rm d}t\, t f^{(\gamma)}(t) = \frac{1}{2} ,
\end{equation}
respectively.

Let us first perform a brief analysis of the general density formula 
(\ref{4.24}), without knowing explicitly $f^{(\gamma)}(t)$.
For odd $\gamma$, the inequality (\ref{4.20}) implies that
$f_{j,N}^{(\gamma)}\le \gamma$ for all $j$, and so $f^{(\gamma)}(t)\le \gamma$
in the whole interval $t\in [ 0,1 ]$.
Consequently,
\begin{equation} \label{4.26}
\tilde{n}(\tilde{x}) \le \frac{\gamma}{2\tilde{x}^2} \left[
1 - (1+2\tilde{x}) {\rm e}^{-2\tilde{x}} \right] \qquad
\mbox{for odd $\gamma$.} 
\end{equation}
In particular, $\tilde{n}(\tilde{x})\le \gamma/(2\tilde{x}^2)$
at large $\tilde{x}$.
This already provides a non-trivial bound. Note that this
relation teaches us that the present analysis, valid for 
integer  values of $\gamma$, cannot be continualized to small 
couplings  to encompass the mean field limit 
$\gamma \to 0$, since at mean-field level, one has $\tilde n \sim \tilde
x^{-2}$.
It is furthermore clear that the asymptotic decay of the particle density
is determined by the behavior of $f^{(\gamma)}(t)$ in the limit $t\to 0$.
Let us assume that $f^{(\gamma)}(t)$ has a power-law behavior
$f^{(\gamma)}(t)\sim_{t\to 0} c t^{\nu}$, where the positiveness and the 
boundedness
of $f^{(\gamma)}(t)$ are ensured by $c>0$ and $\nu\ge 0$, respectively.
The special case of $\nu=0$ corresponds to the situation when
$f^{(\gamma)}(t)$ approaches a positive number as $t\to 0$.
Inserting our assumption into (\ref{4.24}), we get
\begin{equation} \label{4.27}
\tilde{n}(\tilde{x}) \sim (1+\nu)! \frac{c}{2^{1+\nu}} 
\frac{1}{\tilde{x}^{2+\nu}} , \qquad \tilde{x}\to\infty .
\end{equation}
This means that if $\nu>0$ for larger couplings, the asymptotic decay of 
the particle density is faster than the weak-coupling prediction (\ref{2.6}). 

For $\gamma=1$, we have the exact result $f_{j,N}^{(1)}=1$ for all 
$j=0,1,\ldots,N-1$ and particle numbers $N$, so that $f^{(1)}(t)=1$. 
This function fulfils the normalization relations (\ref{4.25}).
The formula (\ref{4.24}) for the density profile becomes
\begin{equation} \label{4.28}
\tilde{n}(\tilde{x}) = \frac{1}{2\tilde{x}^2} \left[
1 - (1+2\tilde{x}) {\rm e}^{-2\tilde{x}} \right] . 
\end{equation}
At large $\tilde{x}$,
\begin{equation} \label{4.29}
\tilde{n}(\tilde{x}) \sim \frac{1}{2\tilde{x}^2}, \qquad
\tilde{x}\to\infty .
\end{equation}
This behavior resembles, up to the renormalization factor $1/2$,
the weak-coupling decay (\ref{2.6}).
The independence of the counter-ion density on the line charge density
$\sigma e$ is present also at the considered finite temperature.

\begin{figure}
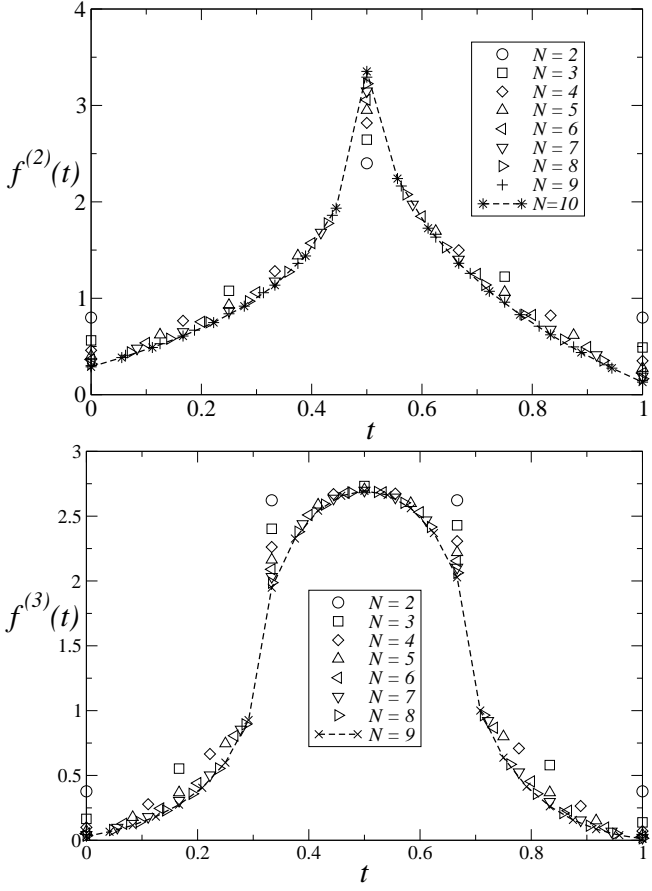

\begin{center}
\includegraphics[width=0.48\textwidth,clip]{Fig4a.eps}
\includegraphics[width=0.48\textwidth,clip]{Fig4b.eps}
\caption{Discrete representations of the functions $f^{(2)}(t)$
$(\Gamma\equiv 2\gamma=4)$ and $f^{(3)}(t)$ $(\Gamma=6)$ for increasing number
of particles $N$.} 
\end{center}
\end{figure}

For $\gamma=2$ and $\gamma=3$ (i.e. $\Gamma=4$ and 6), we performed exact calculations
up to $N=10$ and $N=9$ particles, respectively.
The discrete representations $\{ f_{j,N}^{(2)} \}_{j=0}^{2(N-1)}$ and 
$\{ f_{j,N}^{(3)} \}_{j=0}^{3(N-1)}$ of the corresponding functions 
$f^{(2)}(t)$ and $f^{(3)}(t)$ are presented in Fig. 4.
It is seen that data converge rather quickly when increasing particle numbers.
The continuous functions $f^{(2)}(t)$ and $f^{(3)}(t)$ are well approximated 
by the corresponding discrete plots obtained for $N=10$ and $N=9$ particles 
(dashed lines).
Although the plots look at first sight to be symmetric with respect
to $t=1/2$, they are not.

\begin{figure}[htb]
\begin{center}
\includegraphics[width=0.48\textwidth,clip]{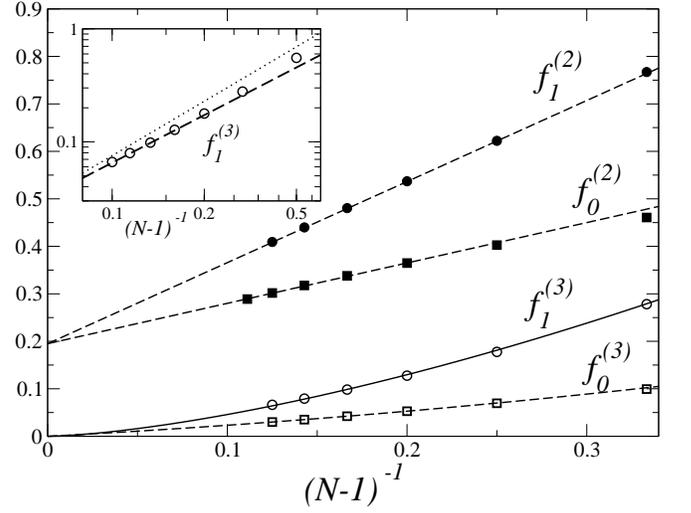}
\caption{The $N$-dependence of discrete sequences 
$f_{0,N}^{(\gamma)}, f_{1,N}^{(\gamma)}$ for $\gamma=2,3$;
their relation to the small-$t$ behavior of the functions 
$f^{(2)}(t)$ and $f^{(3)}(t)$ is explained in the text.
The symbols are for the exact results at finite $N$. 
For $\gamma=\Gamma/2=2$, both $f_1^{(2)}$ and  $f_0^{(2)}$
extrapolate to the same limiting  value (0.195) for $N\to\infty$
(in other words, $\nu=0$ in Eq. (\ref{4.27}).
On the other hand, for $\gamma=3$, a similar analysis provides a vanishing
limit, i.e. $\nu \neq 0$. In the main graph, the dashed lines are 
linear fits while the continuous curve corresponds to a power-law
with exponent 1.45. 
The inset shows $f_1^{(3)}$ on a log-log scale,
to evidence power law behaviour. The dashed line has slope 1.45 while the 
dotted line, given as a guide to the eye, has slope 1.6. 
} 
\end{center}
\end{figure}

In view of the above discussion about the large-distance decay of 
the particle density (\ref{4.27}), the behavior of the functions $f^{(2)}(t)$
and $f^{(3)}(t)$ in the limit $t\to 0$ is of primary importance. 
The first relevant question is whether $f^{(2)}(0)$ and $f^{(3)}(0)$ are
positive or equal to $0$.
Since $t=0$ for $j=0$, we have $f^{(2)}(0) = \lim_{N\to\infty} f_{0,N}^{(2)}$  
and $f^{(3)}(0) = \lim_{N\to\infty} f_{0,N}^{(3)}$.
Given that the ``natural'' variable in the finite-$N$ analysis is $x=1/(N-1)$,
the dependence of the sequences $\{ f_{0,N}^{(2)} \}$ and $\{ f_{0,N}^{(3)} \}$
on $1/(N-1)$ is pictured in Fig. 5.
For $\gamma=2$, the sequence is well fitted by the linear form
$f_{0,N}^{(2)}\sim 0.195+0.85 x$ with $x=1/(N-1)$ (dashed line), i.e.
$f^{(2)}(0)\sim 0.195$ is apparently positive and $\nu=0$ in (\ref{4.27}).
We note that this value is fully corroborated by the analysis of the
behaviour of $f_1^{(2)}$, that is well fitted by 
$f_{1,N}^{(2)}\sim 0.195+1.70 x$ (dashed line in Fig. 5).
We conclude that at $\gamma=2$, the large-distance behavior of the density
exhibits the mean-field behavior $1/\tilde{x}^2$, with a renormalized
prefactor $f^{(2)}(0)/2 \simeq 0.1$, significantly smaller than its
mean-field value of 1, or the $1/2$ value which holds at $\Gamma=2$.
This illustrates enhanced screening at larger Coulombic coupling
$\Gamma$.

For $\gamma=3$, the sequence is well fitted by the quadratic form
$f_{0,N}^{(3)}\sim 0.2 x + 0.32 x^2$ (dashed line), i.e.
$f^{(3)}(0)\sim 0$ and $\nu>0$ in the large distance behavior (\ref{4.27}).
The sequence $\{ f_{0,N}^{(3)} \}$ 
does not provide any information about the value of $\nu$. 
The index $\nu$ can instead be deduced 
from the subsequent sequence $\{ f_{1,N}^{(3)} \}$
with the coordinate $t=1/[3(N-1)]$ which, as $N$ increases, mimics 
the plot of the function $f^{(3)}(t)$ for small $t$.
In the limit $N\to\infty$ $(t=0)$, the sequence $\{ f_{1,N}^{(3)} \}$ 
must converge to the previously obtained $f^{(3)}(0)=0$.  
The sequence is well fitted by a power-law $f_{1,N}^{(3)}\propto x^{\nu}$
with $\nu \simeq 1.45$, 
see the solid line in Fig. 5, indicating that $\nu\simeq 1.45$.
The stability of the fit is documented in the inset,
that shows the sequence $\{ f_{1,N}^{(3)} \}$ on log-log plot;
the dashed line has slope $1.45$ and the dotted line has slope $1.6$. 
A similar value of $\nu$ is obtained from the sequence 
$\{ f_{2,N}^{(3)} \}$, and also from a complementary
direct plot of the data 
of Fig. 3 for $\Gamma=6$ on a log-log scale (not shown).
We therefore see that at $\gamma=3$ the density behaves like 
$\sim 1/\tilde{x}^{a}$ at large distances with an exponent 
$a=2+\nu$ close to 3.5, 
in contrast to the
mean-field prediction.
For this coupling, the asymptotic density decay {\em depends}
on the line charge $\sigma e$. Indeed, returning to
original variables, one has in general
$$
n(x) \propto \sigma^{-\nu} x^{-2-\nu}.
$$

\subsection{Cylinder: Two charged lines}
We now consider the periodic strip of circumference $W$, finite
in the $x$-direction, $x\in [0,d]$, see Fig. 1d.
The equivalent charge densities $\sigma e$ at lines $x=0$ and $x=d$
are neutralized by $N=2\sigma W$ particles of charge $-e$.
The electric field generated by the line charges vanishes,
so that $w(x)=1$ and
\begin{equation} \label{4.30}
w_{\rm ren}(x) = \frac{4\pi}{W^2} 
\exp\left[ \frac{2\pi\gamma}{W} (N-1) x \right] .
\end{equation}
With respect to the equality $d/W = \tilde{d}/(\pi\gamma N)$, 
the interaction strengths (\ref{4.11}) take the form
\begin{equation} \label{4.31}
w_j = \frac{1}{j-\frac{\gamma}{2}(N-1)} \left\{ 1 - 
{\rm e}^{-\frac{4\tilde{d}}{\gamma N}\left[ j-\frac{\gamma}{2}(N-1) \right]} \right\}
\end{equation}
for $j\ne \gamma(N-1)/2$ and $w_j=4\tilde{d}/(\gamma N)$ for $j=\gamma(N-1)/2$.
The density profile is given by
\begin{equation} \label{4.32}
n(x) = \frac{4\pi}{W^2} \sum_{j=0}^{\gamma(N-1)} \langle \Xi_j \Psi_j \rangle
{\rm e}^{-\frac{4\pi}{W}\left[ j-\frac{\gamma}{2}(N-1)\right]x} .
\end{equation}
The reflection symmetry $n(x)=n(d-x)$ implies the relation
\begin{equation} \label{4.33}
w_j \langle \Xi_j\Psi_j \rangle = 
w_{\gamma(N-1)-j} \langle \Xi_{\gamma(N-1)-j}\Psi_{\gamma(N-1)-j} \rangle 
\end{equation}
which is valid for all $j=0,1,\ldots,\gamma(N-1)$.

The continualization of the formula (\ref{4.32}), in the limit $N,W\to\infty$
while keeping the ratio $N/W=2\sigma$, proceeds along the above lines.
As the continuous variable, we choose $t=2[j-\gamma(N-1)/2]/[\gamma(N-1)]$, 
taking values in the interval $[ -1,1]$. 
In the continuum limit, the discrete values 
$f_{j,N}^{(\gamma)}(\tilde{d})=\gamma w_j\langle \Xi_j\Psi_j\rangle$ tend to 
a positive bounded function $f^{(\gamma)}(t,\tilde{d})$.
The continualization of (\ref{4.32}) results in
\begin{equation} \label{4.34}
\tilde{n}(\tilde{x}) = 2 \int_{-1}^1 {\rm d}t\, t f^{(\gamma)}(t,\tilde{d})
\frac{{\rm e}^{-2 t \left(\tilde{x}-\frac{\tilde{d}}{2}\right)}}{
{\rm e}^{t\tilde{d}}-{\rm e}^{-t\tilde{d}}} .  
\end{equation}
The symmetry $n(x)=n(d-x)$ implies
\begin{equation} \label{4.35}
f^{(\gamma)}(t,\tilde{d}) = f^{(\gamma)}(-t,\tilde{d}) ,
\end{equation}
which enables us to rewrite (\ref{4.34}) into a more convenient form
\begin{equation} \label{4.36}
\tilde{n}(\tilde{x}) = 2 \int_0^1 {\rm d}t\, t f^{(\gamma)}(t,\tilde{d})
\frac{\cosh\left[ 2t\left( \tilde{x}-\frac{\tilde{d}}{2}\right)\right]}{
\sinh(t\tilde{d})} .
\end{equation}
The electroneutrality condition (\ref{3.14}) leads to the constraint
\begin{equation} \label{4.37} 
\int_0^1 {\rm d}t\, f^{(\gamma)}(t,\tilde{d}) = 1 .
\end{equation}
According to the contact theorem (\ref{2.1}), the (renormalized) pressure 
is given by 
\begin{equation} \label{4.38}
\tilde{P}=\tilde{n}(0)-1 , 
\end{equation}
i.e.
\begin{equation} \label{4.39}
\tilde{P} = 2 \int_0^1 {\rm d}t\, t \left[ f^{(\gamma)}(t,\tilde{d})
\coth(t\tilde{d}) -1 \right] .  
\end{equation}

For $\Gamma=2$, we have the exact result $f_{j,N}^{(1)}(\tilde{d})=1$
for all $j=0,1,\ldots,N-1$ and particle numbers $N$, which implies 
$f^{(1)}(t,\tilde{d}) = 1$ in the whole interval $t\in [0,1]$.
The pressure
\begin{equation} \label{4.40}
\tilde{P} = 2 \int_0^1 {\rm d}t\, t 
\frac{\exp(-t\tilde{d})}{\sinh(t\tilde{d})} .
\end{equation}
is positive for every distance $d$, so there is always the repulsion between 
two equivalently charged lines at $\Gamma=2$.
For small distances $\tilde{d}$, we have
\begin{equation} \label{4.41}
\tilde{P} = \frac{2}{\tilde{d}} - 1 + \frac{2\tilde{d}}{9}
+ {\cal O}\left( \tilde{d}^2 \right) .
\end{equation}
This expansion resembles the SC one (\ref{3.16}), up to the
renormalization factor $2$ ahead of the $\tilde{d}$ term.
For large distances between the lines, the formula (\ref{4.40}) yields
\begin{equation} \label{4.42} 
\beta P \sim \frac{\pi}{12} \frac{1}{d^2} , \qquad \tilde{d}\to\infty .
\end{equation}
This result coincides, up to a renormalized prefactor, 
with the PB prediction (\ref{2.15}).
The asymptotic decay of the pressure is universal in the sense that
it does not depend on the magnitude of the charge density $\sigma e$ 
on the lines.

It is instructive to compare the exact solution (\ref{4.40}),
obtained in the thermodynamic $N\to\infty$ limit, with the
finite-$N$ results.
For finite $N$, equations (\ref{4.32}) and (\ref{4.38}) imply
\begin{equation} \label{4.43}
\tilde{P} = \frac{8}{\gamma N^2} \sum_{j=0}^{\gamma(N-1)}
\langle \Xi_j \Psi_j \rangle - 1 .
\end{equation}
For $\gamma=1$, we have $\langle \Xi_j\Psi_j \rangle = 1/w_j$
for all $j=0,1,\ldots,N-1$.
To derive the asymptotic $\tilde{d}\to\infty$ behavior of $\tilde{P}$,
we note from (\ref{4.31}) that for $\gamma=1$
\begin{equation} \label{4.44}
\lim_{\tilde{d}\to\infty} \frac{1}{w_j} = \left\{
\begin{array}{ll}
j-\frac{1}{2}(N-1) & \mbox{if $j-\frac{1}{2}(N-1)>0$,} \cr
& \cr 0 & \mbox{if $j-\frac{1}{2}(N-1)\le 0$.}
\end{array}  \right.
\end{equation}
Consequently, for even particle numbers $N$, we have
\begin{equation} \label{4.45}
\lim_{\tilde{d}\to\infty} \tilde{P} = \frac{8}{N^2} \sum_{j=N/2}^{N-1} 
\left[ j - \frac{1}{2}(N-1) \right]- 1 = 0 ,
\end{equation}
while for odd $N$:
\begin{equation} \label{4.46}
\lim_{\tilde{d}\to\infty} \tilde{P} = \frac{8}{N^2} \sum_{j=(N+1)/2}^{N-1} 
\left[ j - \frac{1}{2}(N-1) \right]- 1 = - \frac{1}{N^2} .
\end{equation}
The above results can be understood intuitively by the discreteness
of particles.
If the number of particles is even, i.e. $N=2N^*$, at asymptotically 
large distance each of the charged lines attracts just $N^*$ particles.
The whole system thus consists of two neutral subsystems which do not 
interact with one another.
On the other hand, if the number of particles is odd, $N=2N^*+1$, 
one of the particles is shared by both lines, so that the two double-layers
can never strictly decouple. 
This ``misfit'' particle is responsible for the asymptotic attraction $-1/N^2$
between the charged lines in (\ref{4.46}). 
The asymptotic attraction disappears in the thermodynamic limit $N\to\infty$.
It is interesting that exactly same asymptotic relations (\ref{4.45}) 
and (\ref{4.46}) can be found for higher values of $\gamma$, as was verified
on finite-$N$ calculations for $\gamma=2,3$. 
A detailed discussion about this interesting finite-$N$ phenomenon 
will be published elsewhere \cite{Trizac}. 

\begin{figure}[htb]
\begin{center}
\includegraphics[width=0.48\textwidth,clip]{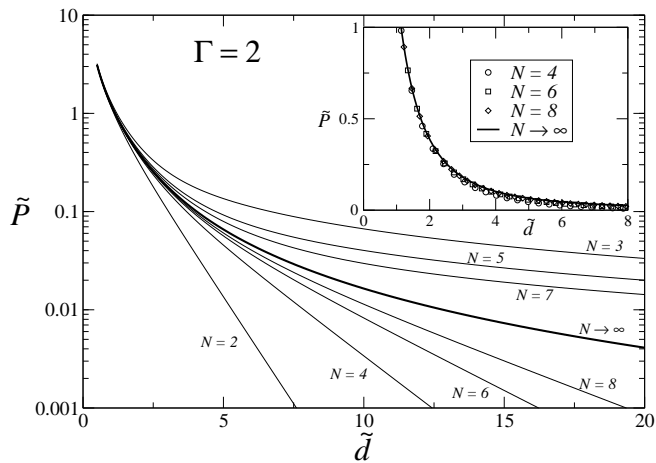}
\caption{Pressure curves for $\Gamma=2$. In the main graph, the
pressure is shifted by $1/N^2$ for odd values of particle numbers $N$.
For the asymptotic $N\to\infty$ result shown by the
thick line, see formula (\ref{4.40}).} 
\end{center}
\end{figure}

The $\Gamma=2$ results for the pressure dependence on the distance between
two lines in the case of finite $N$ are compared with 
the asymptotic $N\to\infty$ result (\ref{4.40}) in Fig. 6.
In the main graph, the pressure is shifted by $1/N^2$ for odd values of 
$N$ (so that it vanishes at large distances) while curves with even $N$ are not shifted.
The decay of the pressure is always monotonous.
We see that curves with even $N$ lie below and the shifted ones 
with odd $N$ lie above
the asymptotic $N\to\infty$ line; the odd and even curves 
systematically ``sandwich'' the asymptotic one.
The inset, where only even $N$ are considered, shows a quick convergence
of the pressure plots for finite particle numbers to the asymptotic one.
To our surprise, relatively small numbers of particles are sufficient
to have realistic estimates of the pressure in the thermodynamic limit,
at least up to ``reasonable'' distances $\tilde{d}\sim 5$. 

\begin{figure}[htb]
\begin{center}
\includegraphics[width=0.48\textwidth,clip]{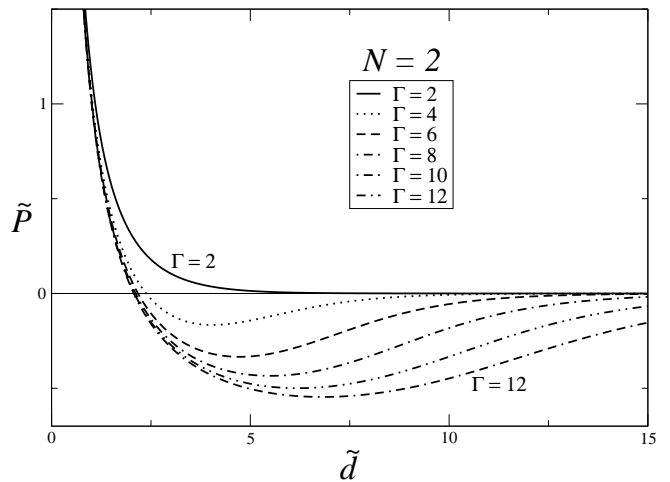}
\caption{Rescaled pressure versus distance between two charged lines 
for $N=2$ particles and the couplings $\Gamma=2,4,\ldots,12$.} 
\end{center}
\end{figure}

The results for the pressure dependence on the distance in the case of 
the smallest particle number $N=2$ and $\Gamma=2,4,\ldots,12$ 
are depicted in Fig. 7.
For $\Gamma=2$, we have the already discussed monotonous decay 
of the pressure to $0$.
For $\Gamma\ge 4$, the pressure becomes negative at a $\Gamma$-dependent
distance and remains negative up to $\tilde{d}\to\infty$, i.e. 
there is no further intersection of the curve with the $\tilde{P}=0$ axis.
We learn that the attraction between equivalently charged lines is not
associated only with the thermodynamic limit, but manifests itself
(at sufficiently large $\Gamma$) even for $N=2$ particles.  

Pressure curves for $\Gamma=4$ and $\Gamma=6$ in Fig. 8 are presented 
separately for even $N=2,4,8$ even (the main graph) and odd 
$N=3,5,7$ (the inset).
We see that by increasing $N$ the difference between plots becomes very
small and so they are probably very close to the asymptotic line,
at least for $\tilde{d}\le 5$.
As in the $N=2$ case, the pressure crosses the $\tilde{P}=0$ line
at some distance. 
This distance, evaluated at $N=8$, is indicated for $\Gamma=4,6$ 
by open circles in Fig. 3; we see a good agreement with 
the SC phase diagram (solid line).
On the other hand, the distance of the maximum attraction (open squares) 
is relatively far away from the corresponding dashed line; this might be
caused by an extended plateau around the minimum point. 
As before, after crossing the $\tilde{P}=0$ line, the curves remain
in the attraction region up to $\tilde{d}\to\infty$.
This fact sheds doubts on the existence of the upper branch of
the phase diagram (for $N\to\infty$) in Fig. 3.    
We cannot answer this question because the finite-$N$ calculations
do not reflect adequately the pressure in the thermodynamic limit just 
in the region of large $\tilde{d}$.

\begin{figure}[htb]
\begin{center}
\includegraphics[width=0.48\textwidth,clip]{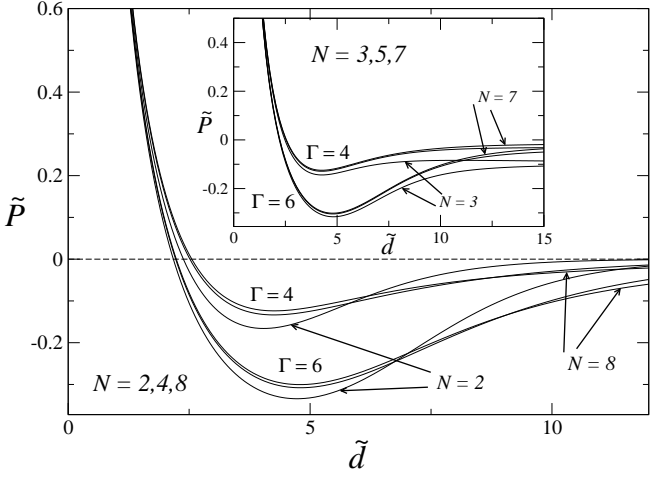}
\caption{Pressure curves for $\Gamma=4$ and $\Gamma=6$.
Results are presented separately for even $N=2,4,8$ (the main graph) and 
odd $N=3,5,7$ (the inset).} 
\end{center}
\end{figure} 

We end up this section by an analysis of the small-distance expansion
of the pressure; see formula (\ref{2.14}) for the weak-coupling regime,
(\ref{3.16}) for the SC regime and (\ref{4.41}) for $\Gamma=2$.
The leading small-distance term is always $2/\tilde{d}$ as
a consequence of the spatial homogeneity of the particle density 
$\tilde{n}(\tilde{x})\sim 2/\tilde{d}$ in the limit $\tilde{d}\to 0$. 
The next (constant) term is equal to $-1/3$ in the weak-coupling limit
and to $-1$ in the SC limit and for $\Gamma=2$.
The value $-1$ was detected also in finite-$N$ calculations for
$\Gamma=4,6$ --as Fig. 9 shows-- 
and is expected to persist up to $\Gamma\to\infty$.
From the point of view of the contact theorem (\ref{4.38}), this means
that in the limit $\Gamma\to 0$ the contact density behaves like
$\tilde{n}(0)\sim 2/\tilde{d}+ 2/3 + {\cal O}(\tilde{d})$, while
for $\Gamma=2,4,6,\ldots$ it behaves like
$\tilde{n}(0)\sim 2/\tilde{d}+{\cal O}(\tilde{d})$. 
It is not clear at which $\Gamma<2$ the fundamental change in the
short-distance behavior of $\tilde{n}(0)$ starts; in the subsequent
analysis, 
we shall restrict ourselves to the region $\Gamma\ge 2$, where
\begin{equation} \label{4.47}
\tilde{P} = \frac{2}{\tilde{d}} - 1 + A_N(\Gamma) \tilde{d} 
+ {\cal O}(\tilde{d}^2) .
\end{equation}
We have at our disposal the exact information about the $\Gamma\to\infty$
limit of the prefactor: 
\begin{equation} \label{4.48}
\lim_{\Gamma\to\infty}  \frac{9 \Gamma}{2} A_{\infty}(\Gamma) =  1.
\end{equation}
The simplest Pad\'{e} approximant corresponding to this asymptotic 
formula is
\begin{equation} \label{4.49}
A_{\infty}(\Gamma) = \frac{2}{9 \Gamma} \frac{\Gamma+a}{\Gamma+b} ,
\end{equation}
where the free coefficients $a,b$ should be fixed by the known 
(exact or approximate) values of $A_{\infty}(\Gamma)$ at some $\Gamma$-points.
We know the exact result at $\Gamma=2$: $A_{\infty}(\Gamma=2) = 2/9$.
We were able to evaluate accurately $A_{\infty}(\Gamma)$ for $\Gamma=4,6$;
the analysis for $\Gamma=6$ is presented in Fig. 9.
Considering $\tilde{P}-2/\tilde{d}+1$ as the function of $\tilde{d}$
in the main graph, $A_N$ is nothing but the slope $y/x$ of the 
$N=6,7,8$ data curves, taken in the small-$x$ limit.
In the inset, the slopes $A_N$ $(N=6,7,8)$ are fitted against $1/N$
with the function $y=0.0454-0.013 x$, hence the value 
$A_{\infty}(\Gamma=6)\simeq 0.045$ for large $N$.
The same scenario applies to $\Gamma=4$, where the limiting
slope is $A_{\infty}(\Gamma=4) \simeq 0.075$.
The three results at $\Gamma=2,4,6$ are perfectly matched by
the two-parameter Pad\'{e} approximant (\ref{4.49}) if we choose 
$a=1/4$ and $b=-7/8$, i.e.
\begin{equation} \label{4.50}
A_{\infty}(\Gamma) = \frac{2}{9 \Gamma} \frac{\Gamma+1/4}{\Gamma-7/8} ,
\qquad \Gamma\ge 2 .
\end{equation}
It is quite remarkable that this simple form accounts for the
exactly know results at $\Gamma=2$ and $\Gamma \to \infty$,
together with the numerically accurate data in the thermodynamic
limit obtained at $\Gamma=4$ and $\Gamma=6$. It is however 
inapplicable to the mean-field limit $\Gamma\to 0$.

\begin{figure}[htb]
\begin{center}
\includegraphics[width=0.48\textwidth,clip]{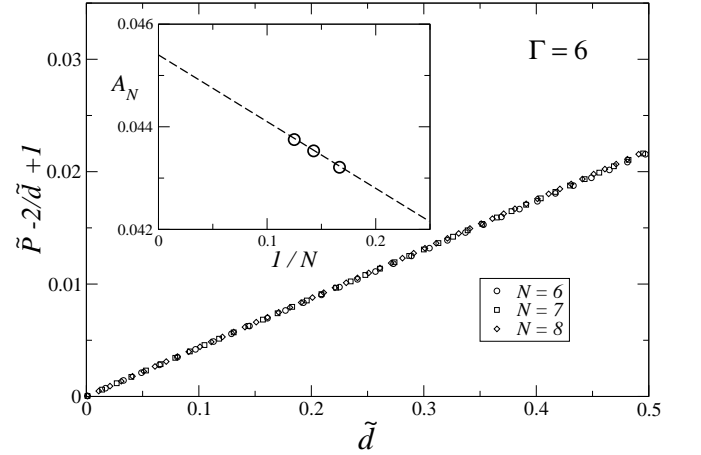}
\caption{The procedure of fitting the coefficient $A_{\infty}(\Gamma=6)$, 
defined by the expansion (\ref{4.47}). $A_N$ corresponds to the slopes
of the $N=6,7,8$ data sets in the main graph. In the inset, $A_N$
are fitted linearly against $1/N$ to obtain $A_{\infty}(\Gamma=6)\sim 0.045$.
Note the vertical scale in the inset that indicates that the thermodynamic
limit is already closely approached by systems with $N=6$ particles.
Very similar results were found at $\Gamma=4$, with different numerical
constants.} 
\end{center}
\end{figure} 

The result (\ref{4.50}), when substituted into the expansion (\ref{4.47}), 
can be used to improve the phase diagram in Fig. 3, that
follows from the SC equation of state (\ref{3.16}).
The problem of our original SC phase diagram of Fig. 3 is that 
it includes $\Gamma=2$ in the attractive regime (for some distances
$\tilde d$ around 4), while we have shown that for $N\to \infty$,
there is no attraction between lines at the exactly
solvable $\Gamma=2$ case (see e.g. Figure 6).
It can be seen in Fig. 3 that making use of the improved equation of state 
(\ref{4.47}) together with (\ref{4.50}),
leads to a
shift of the attraction border towards higher couplings (see the dashed-dotted
line). More precisely, the critical $\Gamma$ below which no attraction
is possible is changed from a value 1.77 with (\ref{3.16}), to 2.81 
with (\ref{4.50}). Since our exact results indicate the possibility
of attraction at $\Gamma=4$ and 6, but none at $\Gamma=2$, the second estimation 
seems more reliable.

\renewcommand{\theequation}{5.\arabic{equation}}
\setcounter{equation}{0}

\section{Conclusion}
In this paper, we have studied 2D models of counter-ions at and between
charged lines. Since only one type of counter-ions was 
considered (no salt), these ions can be considered as point-like
without any ensuing pathology.
2D models, while maintaining the essence of 3D Coulomb models,
are simpler to handle analytically as well as numerically.
In the one-line geometry, we focused on the large-distance decay
of the particle density.
In the two-line geometry, the small distance behavior of
the pressure was the center of interest (although of particular
significance, the large distance behaviour is more difficult to
obtain, and could only be addressed in particular cases).
The possibility of an attraction (negative pressure) for some distances
between equivalently charged lines, mediated by counterions, 
was investigated.   

The weak-coupling limit has basically the same mean-field 
Poisson-Boltzmann (PB) form in 
any dimension.
For one-line geometry, the counter-ion density falls at asymptotically
large distances like the inverse-power-law $1/x^2$ which does not depend
on the magnitude of the line charge $\sigma e$. 
The same phenomenon is observed in the two-line geometry: 
The asymptotic decay of the pressure does not depend on $\sigma e$.
This phenomenon occurs in both 2D and 3D.
The pressure is always positive: the attraction phenomenon is
absent for weak couplings.

The strong coupling (SC) analysis presented here is the 2D adaptation
of the method \cite{Samaj10} based on the harmonic expansion of
the interaction energy around the ground-state Wigner crystal
formed by counter-ions.
The method is applicable to small distances.
The small-distance expansion of the pressure between two lines provides 
a  phase diagram --in Fig. 3-- which includes the attraction
region. Our strong-coupling expansion differs from those
that have previously been proposed: it allows to compute the
corrections to the leading order term, which is not the case of 
the original
method of Netz and collaborators \cite{Boroudjerdi05,Moreira00,Netz01},
that nevertheless successfully predicts the leading term.
Our approach also differs from that of Refs. \cite{Lau1,Lau2},
where the excitations considered (counter-ions displacements restricted to
the charged interface) are not those that turn relevant at short
distances. As a consequence, the predictions of Refs. \cite{Lau1,Lau2}
do not cover the short-range phenomena that we have studied here
under strong coupling.

The intermediary (i.e. between weak- and strong-coupling regimes)
values of $\Gamma=2,4,6$ are studied by using a previously developed
Grassmann variables formalism \cite{Samaj95,Samaj04a,Samaj04b}. 
As a constraining domain for counter-ions, we choose the surface
of a cylinder; this enables us to mimic infinite systems by
finite ones containing finite numbers of particles $N$.
We systematically observed that a system with as little as
$N=5$ to 10 particles may be considered as ``large'',
in that it is already close to the thermodynamic limit.

The case $\Gamma=2$ is solved exactly, for finite as well as infinite $N$.
It shares many features with the PB theory: The particle density decay in
$1/x^2$ does not depend on $\sigma e$, the pressure is always repulsive
and its large-distance asymptotic is independent of $\sigma e$.

The couplings $\Gamma=4,6$ are investigated for finite particle numbers $N$.
In the one-line problem, the particle density decay is still of the PB type
$1/x^2$ for $\Gamma=4$.
For $\Gamma=6$, there are strong indications that the counter-ion density
behaves like $1/x^{a}$ for large $x$ where the exponent $a$
is close to 1.5 (we found $1.4<a<1.6$). 
This signals the breaking down of the large-distance PB theory and the
dependence of the asymptotic density on $\sigma e$.
It has been argued on the contrary that a strongly coupled
double-layer behaves, at large distances where counter-ions correlations
should be less important, as predicted by a suitably renormalized mean-field 
approach \cite{Shklovskii,Levin09}. The data reported here provide evidence, for two dimensional
systems, that this is not always the case. The corresponding
question for 3D systems that are the main
objects of interest in Refs \cite{Shklovskii,Levin09} (i.e. 
with $1/r$ interactions instead of $\log r$)
remains open.
As concerns the pressure between equivalently charged lines,
it is highly non-trivial even for $N=2$ particles (see Fig. 7).
The monotonous decay at $\Gamma=2$ changes for $\Gamma\ge 4$ 
and turns into a profile
with an attractive regime, starting from a certain distance. 
Increasing $N$, the results converge quickly for intermediate distances
of interest.
We see that the attraction phenomenon is not restricted to the
thermodynamic limit, but takes place for small particle numbers and
relatively small values of $\Gamma$ (4 and 6).
The finite-$N$ analysis of the equation of state enabled us to
improve the phase diagram evaluated in the SC regime. 
 
Many questions are still open. Among interesting perspectives is 
the question the effective interaction between arbitrarily 
shaped objects. Another relevant problem lies in the generalization 
of the present ideas and methods to system where not only one type 
of micro-ions is present, such as electrolytes.

\begin{acknowledgments}
L. \v{S}. is grateful to LPTMS for hospitality.
The support received from Grant VEGA No. 2/0113/2009 and 
CE-SAS QUTE is acknowledged. 
\end{acknowledgments}


\begin{thebibliography}{1}


\bibitem{Hunter}
R.J. Hunter, Foundations of Colloid Science,
Oxford University Press (2005).

\bibitem{GC} 
G.L. Gouy, J. de Phys. {\bf 9} 457 (1910);
D.L. Chapman, Phil. Mag. {\bf 25} 475 (1913).

\bibitem{Attard}
P. Attard, 
Adv. Chem. Phys. XCII, 1 (1996).

\bibitem{JPH}
J.P. Hansen and H. L\"owen, 
Annu. Rev. Phys. Chem. {\bf 51}, 209 (2000).

\bibitem{Messina09}
R. Messina, J. Phys.: Condens. Matter {\bf 21}  113102 (2009).

\bibitem{Boroudjerdi05}
H. Boroudjerdi, Y.-W. Kim, A. Naji, R. R. Netz, X. Schlagberger, A. Serr,
 Phys. Rep. {\bf 416}, 129 (2005).



\bibitem{Andelman06}{D. Andelman, in {\it Soft Condensed Matter Physics in 
Molecular and Cell Biology}, edited by W. C. K. Poon and D. Andelman
(Taylor \& Francis, New York, 2006), Chapt. 6.}


\bibitem{Attard88}{P. Attard, D. J. Mitchell and B. W. Ninham,
J. Chem. Phys. {\bf 88}, 4987 (1988); {\it ibid} {\bf 89}, 4358 (1988).}

\bibitem{Podgornik88} 
R. Podgornik and B. Zeks,
J. Chem. Soc. Faraday Trans. {\bf 4}, 611 (1988).

\bibitem{Podgornik90}{R. Podgornik, J. Phys. A {\bf 23}, 275 (1990).}

\bibitem{Netz00}{R. R. Netz and H. Orland, 
Eur. Phys. J. E {\bf 1}, 203 (2000).}


\bibitem{Rouzina96}{I. Rouzina and V. A. Bloomfield, 
J. Phys. Chem. {\bf 100}, 9977 (1996).}

\bibitem{Grosberg02}{A. Y. Grosberg, T. T. Nguyen and B. I. Shklovskii,
Rev. Mod. Phys. {\bf 74}, 329 (2002).}

\bibitem{Levin02}{Y. Levin, Rep. Prog. Phys. {\bf 65}, 1577 (2002).}

\bibitem{Moreira00}{A. G. Moreira and R. R. Netz, Europhys. Lett. {\bf 52}, 
705 (2000).}

\bibitem{Netz01}{R. R. Netz, Eur. Phys. J. E {\bf 5}, 557 (2001).}

\bibitem{Moreira}{A. G. Moreira and R. R. Netz, Phys. Rev. Lett. {\bf 87}, 
078301 (2001); Eur. Phys. J. E {\bf 8}, 33 (2002).}

\bibitem{Lau1}
A.W.C Lau, D. Levine and P. Pincus, 
Phys. Rev. Lett. {\bf 84}, 4116 (2000).

\bibitem{Lau2}
A.W.C Lau, P. Pincus, D. Levine, H.A. Fertig, 
Phys. Rev. E {\bf 63}, 051604 (2001).

\bibitem{Chen06}{Y. G. Chen and J. D. Weeks, 
Proc. Natl. Acad. Sci. U.S.A. {\bf 103}, 7560 (2006);
J.M. Rodgers, C. Kaur, Y.G. Chen and J.D. Weeks, Phys. Rev. Lett. 
{\bf 97}, 097801 (2006). } 

\bibitem{Santangelo06}{C. D. Santangelo, Phys. Rev. E {\bf 73}, 041512 (2006).}

\bibitem{Jho08}
Y.S. Jho, M. Kanduc, A. Naji, R. Podgornik, M.W. Kim and P.A. Pincus,
Phys. Rev. Lett. {\bf 101}, 188101 (2008).

\bibitem{Dean09}
D.S. Dean, R.R. Horgan, A. Naji, R. Podgornik,
J. Chem. Phys. {\bf 130}, 094504 (2009).

\bibitem{Hatlo10}{M. Hatlo and L. Lue, 
Europhys. Lett. {\bf 89}, 25002 (2010).}

\bibitem{Kanduc1}
M. Kanduc, M. Trulsson, A. Naji, Y. Burak, J. Forsman, R. Podgornik,
Phys. Rev. E {\bf 78}, 061105 (2008).
 
\bibitem{Kanduc2}
M. Kanduc, A. Naji, J. Forman, R. Podgornik,
J. Chem. Phys. {\bf 132}, 124701 (2010); 

 
\bibitem{Samaj10}{L. \v{S}amaj and E. Trizac, 
{\it Strong-Coupling Theory of Counter-Ions at Charged Plates}, 
submitted (2010), arXiv:1009.4640}


\bibitem{Naji05}{A. Naji and R. R. Netz, Phys. Rev. Lett. {\bf 95}, 185703
(2005); Phys. Rev. E {\bf 73}, 056105 (2006).}

\bibitem{Burak06}{Y. Burak and H. Orland, 
Phys. Rev. E {\bf 73}, 010501(R) (2006).}

\bibitem{Choquard81}{Ph. Choquard, Helv. Phys. Acta {\bf 54}, 332 (1981).}

\bibitem{Jancovici92}{B. Jancovici, in {\it Inhomogeneous Fluids}, 
edited by D. Henderson (Dekker, New York, 1992), pp. 201-237.}

\bibitem{Samaj95}{L. \v{S}amaj and J. K. Percus, 
J. Stat. Phys. {\bf 80}, 811 (1995).}

\bibitem{Samaj04a}{L. \v{S}amaj, J. Stat. Phys. {\bf 117}, 131 (2004).}

\bibitem{Samaj04b}{L. \v{S}amaj, J. Wagner and P. Kalinay, 
J. Stat. Phys. {\bf 117}, 159 (2004).}

\bibitem{Shklovskii}
B.I. Shklovskii,
Phys. Rev. E {\bf 60}, 5802 (1999).

\bibitem{Levin09}
A.P. dos Santos, A. Diehl and Y. Levin,
J. Chem. Phys. {\bf 130}, 124110 (2009).


\bibitem{Henderson}{
D. Henderson and L. Blum, J. Chem. Phys. {\bf 69}, 5441 (1978);
D. Henderson, L. Blum and J. L. Lebowitz, 
J. Electroanal. Chem. {\bf 102}, 315 (1979);
D. Henderson and L. Blum, J. Chem. Phys. {\bf 75}, 2025 (1991).}

\bibitem{Carnie81}{S. L. Carnie and D. Y. C. Chan, 
J. Chem. Phys. {\bf 74}, 1293 (1981).}

\bibitem{Wennerstrom82}{H. Wennerstr\"om, B. J\"onsson and P. Linse,
J. Chem. Phys. {\bf 76}, 4665 (1982).}

\bibitem{TT03}
G. T\'ellez and E. Trizac
J. Chem. Phys. {\bf 118}, 3362 (2003).

\bibitem{ET2000}
E. Trizac, Phys. Rev. E {\bf 62}, R1465 (2000).

\bibitem{Gradshteyn}{I. S. Gradshteyn and I. M. Ryzhik, 
{\it Table of Integrals, Series and Products}, 5th edn.
(Academic Press, London, 1994).}

\bibitem{Gulbrand84}{L. Gulbrand, B. J\"onson, H. Wennerstr\"om and P. Linse,
J. Chem. Phys. {\bf 80}, 2221 (1984).}

\bibitem{Kjellander84}{R. Kjellander and S. Mar\v{c}elja,
Chem. Phys. Lett. {\bf 112}, 49 (1984).} 

\bibitem{Kekicheff93}{P. K\'{e}kicheff, S. Mar\v{c}elja, T. J. Senden and
V. E. Shubin, J. Chem. Phys. {\bf 99}, 6098 (1993).} 

\bibitem{Berezin66}{F. A. Berezin, {\it The Method of Second Quantization}
(Academic Press, New York, 1966).}

\bibitem{Trizac}{E. Trizac and G. Tell\'{e}z, to be published.}









\end{thebibliography}
\end{document}